\documentclass[aps,prper,reprint,groupedaddress]{revtex4-2}

\usepackage[utf8]{inputenc}
\usepackage{microtype}
\usepackage{placeins}

\usepackage{amsmath,amssymb}
\usepackage{nicefrac}
\usepackage{braket}

\usepackage{graphicx}
\usepackage{booktabs}
\usepackage{tabularx}
\usepackage{makecell}
\usepackage[caption=false]{subfig}
\usepackage{rotating} % add in preamble
\usepackage{enumitem}
\usepackage[normalem]{ulem}
\usepackage{csquotes}
\usepackage{eurosym}
\usepackage{todonotes}

\PassOptionsToPackage{dvipsnames}{xcolor} % <-- must be BEFORE any xcolor is loaded
\usepackage{xcolor}
\usepackage[hidelinks]{hyperref}

\usepackage{url}

\providecommand{\urlprefix}{URL }

\makeatletter
\providecommand{\href@noop}[2]{#2}
\makeatother

\begin{document}

\title{AI-based Verbal and Visual Scaffolding in a Serious Game: Effects on Learning and Cognitive Load}

% --- Authors + affiliations (RevTeX style; no $^1$ etc.) ---

% ----- Authors (manual superscripts like before) -----

\author{Caroline Wermann$^{1}$}
\author{Karina E. Avila$^{1}$}
\author{Sebastian Andr\'e$^{2}$}
\author{Julia C. Draeger$^{3}$}
\author{Alvar Goetze$^{1}$}
\author{Jochen Kuhn$^{1}$}
\author{Maite Maurer$^{4}$}
\author{Sascha Mehlhase$^{4}$}
\author{Nikola Merkas$^{6}$}
\author{Fabian Schrodt$^{7}$}
\author{Stefan K\"uchemann$^{1}$}
\email{s.kuechemann@lmu.de}

% ----- Numbered affiliations (manual list like before) -----

\affiliation{$^{1}$Faculty of Physics, Ludwig-Maximilians-University Munich, Geschwister-Scholl-Platz 1, 80539 Munich, Germany}
\affiliation{$^{2}$Munich Center for Quantum Science and Technology (MCQST), Schellingstr. 4, 80799 Munich, Germany}
\affiliation{$^{3}$Department of Chemistry of the TUM School of Natural Science, Technical University of Munich, Lichtenbergstr. 4, 85748 Garching, Germany}
\affiliation{$^{4}$Munich Quantum Valley (MQV), Leopoldstr. 244, 80807 Munich, Germany}

\affiliation{$^{6}$Studio Merkas, Immenhoferstr. 17, 70180 Stuttgart, Germany}

\affiliation{$^{7}$Quantum Gaming GmbH, Panoramastr. 71, 72070 Tuebingen, Germany}

% If needed, attach Quantum Gaming GmbH to the correct author(s) like:
% \author{<Name>}
% \affiliation{Quantum Gaming GmbH, Panoramastr. 71, 72070 Tuebingen, Germany}

\begin{abstract}
Due to their interactive nature, serious games offer valuable opportunities for supporting learning in educational contexts. Recent advances in large language models (LLMs) have further opened the door to new forms of personalized scaffolding in education. In this study, we combine both worlds and study three types of AI-based scaffolding designs in a serious game: (i) no scaffolding, (ii) chat-based (verbal) scaffolding provided by an AI-based non-player character (NPC), and (iii) combined chat-(verbal) and action-based (visual) scaffolding in which the AI may both try to explain or demonstrate the next step towards a solution. The scaffolding conditions are embedded in Qookies, a serious game designed to introduce fundamental concepts of quantum technologies. A total of 152 school students, university students, and members of the general public were randomly assigned to one of the three conditions. The results show that all groups experience significant learning gains, confirming the overall effectiveness of the serious game itself. No significant differences in learning outcomes emerged between scaffolding conditions. However, intrinsic cognitive load was lower in the combined chat-and-action (verbal+visual) scaffolding condition compared to the chat (verbal)-only condition, suggesting that visual demonstrations may offer more accessible support. Interaction analyses further revealed that players engaged with the AI character primarily for level-related questions and action recommendations, while deeper interactions were relatively rare.
\end{abstract}

\keywords{Game-based learning; Serious game; Quantum technologies; Verbal scaffolding; Visual scaffolding}

\maketitle

% ============================================================
% Your main text starts here
% ============================================================

\section{Introduction}
\label{sec:Intro}

Serious games have been widely discussed in educational research as a promising approach to support learning~\cite{wouters2013,mayer2014b,rodriguez2024educational,Schrader2022}. Their interactive nature allows learners to actively engage with concepts, manipulate variables, and explore cause–effect relationships in ways that are difficult to realize through static instructional media~\cite{rutten2012learning,moreno2007}. When grounded in established learning theories, serious games can support conceptual understanding, foster motivation, and promote inquiry-oriented learning processes~\cite{Schrader2022,wouters2013,Mayer2014}.

Furthermore, a recent international comparison highlights that in countries of the Organization for Economic Co-operation and Development (OECD), a large majority of adolescents regularly use digital devices to play video games and that by the age of 15, nearly all learners have access to mobile devices and digital games~\cite{OECD2025}. Beyond indicating high levels of exposure, this widespread engagement suggests that digital games constitute a familiar and accessible medium for many learners. From an educational perspective, this creates an opportunity to build on existing practices and experiences by embedding instructional goals within game-based environments.

%Serious games, which purposefully combine gameplay with instructional objectives, can leverage this familiarity to support learning about complex subject matter. When designed in alignment with educational theory, they offer opportunities to present abstract content in interactive, motivating, and age-appropriate ways, while embedding guidance and feedback directly into the learning environment.

However, the effective integration of serious games into education is not straightforward. Research has shown that the mere use of games does not automatically lead to improved learning outcomes; rather, learning gains depend on how games are designed and integrated into instruction \cite{wouters2013,Schrader2022,iten2016learning,keller2021potential}. Therefore, a central challenge in the design of game-based learning environments is how to provide appropriate instructional support without compromising opportunities for active problem solving~\cite{de1998scientific,Schrader2022}.

Scaffolding offers one approach to addressing this challenge. It refers to temporary support that helps learners build on their prior knowledge and enables them to engage with tasks that would otherwise exceed their current abilities \cite{wood1976,vygotsky1978}. In digital learning environments, such support can take various forms, including task decomposition, cueing, or self-reflection questions, and must be carefully aligned with learners’ cognitive resources and prior knowledge~\cite{Sweller1998}. In this way, scaffolding can affect students' cognitive load and learning outcomes \cite{Nooijen2024}.

Recent developments in large language models (LLMs) introduce new possibilities for implementing adaptive scaffolding in digital learning environments~\cite{steinert2024harnessing, ma2025dbox}. In serious games, AI-based non-player characters (NPCs), for example, can provide personalized explanations, hints, or demonstrations that respond to learners’ actions and questions~\cite{park2023generative,alavi2024mcpdial,ngu2025generative}. Such agents have the potential to individualize support and regulate task demands dynamically. At the same time, their educational value cannot be assumed a priori, especially since it is known that AI-generated feedback can introduce risks such as hallucinations—instances where the system produces incorrect or fabricated information—potentially undermining learning or increasing cognitive load if not carefully managed~\cite{kasneci2023chatgpt,kuchemann2025opportunities,krupp2024unreflected}. At the same time, it is critical that the learning support from LLMs may impair students' cognitive activity, potentially leading to an unreflected acceptance of the LLMs' output and low retention \cite{krupp2024unreflected,kosmyna2025your}. Whether AI-based support contributes to learning depends on how it is embedded in the instructional design, how learners interact with it, and how it affects cognitive load and engagement.

In this paper, we demonstrate effects of different types of personalized scaffolding via an AI-based NPC in a theory-based serious game, named Qookies. Specifically, we compare three modes of scaffolding: (i) no scaffolding, (ii) generative AI-based verbal scaffolding, and (iii) a combination of generative AI-based verbal scaffolding and an AI-based visual scaffolding.

\section{Theoretical background}
\label{sec:Theory}

\subsection{Game-based Learning}
\label{subsec:GameBasedLearning}
Game-based learning is the process and practice of learning by using games. It is a type of gameplay with defined learning outcomes~\cite{shaffer2005}. There is an ongoing debate among scholars about the definition of a game. A widely adopted definition, proposed by~\cite{tekinbas2003}, describes a game as ``a system in which players engage in an artificial conflict, defined by rules, that results in a quantifiable outcome''. 

Unlike gamification, where game elements such as point systems, awards or leaderboards are used in non-game contexts~\cite[]{deterding2011}, serious games are fully developed games. They leverage the positive characteristics of games while serving a specific, non-entertainment purpose~\cite{ritterfeld2009}. They differ from recreational games due to the use of learning theories and learning objectives that guide the game design. A subgroup of serious games are educational games, that are specifically designed to support academic learning~\cite{mayer2014b}. 

Several studies show that game-based learning can have a positive impact on motivation~\cite{bawa2018, chang2017, wouters2013}, learning effectiveness~\cite{bawa2018, chang2017, wouters2013}, and academic performance~\cite{karakocc2022}. Serious games are already frequently used to teach both classical and quantum physics~\cite{seskir2022, piispanen2025}.

\subsection{Scaffolding Theory}
\label{subsec:ScaffTheory}
Scaffolding is the process by which a tutor or an experienced person helps a learner to complete a task that they currently cannot solve on their own. The more challenging parts of the task are simplified or gradually controlled, allowing the learner to focus on aspects within their competence. The level of support is adapted to suit the learner's cognitive potential. As the learner becomes more proficient, support is gradually reduced until they can complete the task independently. This definition from~\cite{wood1976} is grounded in~\citeauthor{vygotsky1978}'s concept of the zone of proximal development (ZPD), which describes the space between what learners can achieve alone and what remains beyond reach even with guidance. 

Scaffolding can be understood as a means of regulating task complexity in a way that makes otherwise inaccessible problem-solving steps attainable. To understand the underlying mechanism of how scaffolding supports learning, it is necessary to examine how scaffolding interacts with the limits of learners' working memory and schema construction.

\subsection{Cognitive Load Theory}
\label{subsec:CLTheory}
The Cognitive Load Theory (CLT) describes how the working memory affects learning~\cite{Sweller1988, Sweller1989, sweller2023}. It posits that the working memory can only hold a limited amount of elements for a limited time, while the long-term memory has effectively unlimited storage. Thus, the working memory represents the bottleneck in the process of knowledge formation. Cognitive load is the mental effort required to process information. When the amount of information exceeds the capacity of the working memory, that is, when the cognitive load is too high, information cannot be maintained and is lost from working memory. \cite{Sweller1988, Sweller1989, sweller2023} distinguishes three types of cognitive load: the intrinsic cognitive load (ICL), which is determined by the inherent difficulty of the subject matter and the prior knowledge of the learner; extraneous cognitive load (ECL), which is the unnecessary burden, influenced by the presentation and design of the learning materials; germane cognitive load (GCL), which is associated with processes, such as the development of schemata, and the integration of knowledge into long-term memory~\cite{Sweller1998}. 

For a given task and a learner’s specific prior knowledge, ICL remains fixed. It can only be changed by modifying the task demands or learners' knowledge level. It is determined by the task’s element interactivity, the number of information elements that must be processed simultaneously to achieve understanding, with each element representing a unit that must be comprehended and learned~\cite{Sweller2010}.

Element interactivity is also essential for ECL. Poor instructional design can increase unnecessary interactivity and thus contribute to ECL~\cite{Sweller2010}. Scaffolding techniques, such as cueing or separating, can be used to reduce avoidable levels of element interactivity by directing learners' attention or combining information into sets that are processed as a whole, respectively~\cite{Nooijen2024}. 

%Participants in Condition~1 have access to a chat-AI. The NPC can explain the potential solution steps and thereby direct the players' attention. However, the cognitive effort of recognizing the described objects and placing them in the correct location remains with the learners. If players do not immediately recognize which objects are intended, this can lead to the split-attention effect, which means additional search and integration tasks~\cite{chandler1992}. This increases the ECL compared to scaffolding, where the required objects are immediately apparent. 

%The action-AI available to participants in Condition~2 allows the NPC to perform that action itself. By attempting a solution when prompted, it collects and combines the necessary objects, directing the participants' attention immediately to the truly relevant objects. The additional step of transfer is eliminated, further reducing the ECL. 

\subsection{ICAP framework}
The ICAP framework categorizes cognitive engagement into four modes: Interactive, Constructive, Active, and Passive. It posits that as students' engagement with learning materials increases from Passive to Interactive, their learning outcomes improve~\cite{chi2014}. 

The passive mode of engagement is defined as ``receiving information without doing anything else learning related''~\cite{chi2014}. To demonstrate active engagement with the learning material, learners must perform a motoric action that focuses their attention on the subject. If learners generate additional outputs that contain new ideas that go beyond what was in the learning material, they engage in constructive behavior. \cite{chi2018} define interactive behavior as an interaction between two peers, mainly as dialogues, in which the contributions must be primarily constructive, and the speakers take turns in the conversation.

%The ICAP framework serves as the basis for categorizing player-AI communication. The written exchange between  players and the NPC is evaluated in an exploratory manner in order to assess the players' usage behavior of the chat function and the degree of their engagement with the game content.

\subsection{Human-AI-collaboration}
Pedagogical agents play an increasingly important role in the design of learning environments for self-regulated learning. Pedagogical agents are intelligent dialogue systems that communicate with their users via natural language~\cite{khosrawi2023} and support them through personalized explanations, encouragement of reflection processes, feedback, and problem-solving strategies~\cite{gubareva2020, wollny2021}. 
Research on pedagogical agents distinguishes between different roles that differ significantly in their function and effect on learning processes. Traditionally, such agents are designed as tutors who support learners through scaffolding, i.e., providing guidance and support within the learner’s ZPD~\cite{vygotsky1978, wood1976}. These tutor agents are particularly effective in imparting factual knowledge and in clearly structured learning environments as they provide orientation and reduce cognitive load~\cite{schroeder2013}. 

In the last decade, the role of these agents has changed from purely instructional and tutoring roles to adaptive and context-sensitive behavior. Since the advent of powerful AI systems, these agents have been increasingly used for collaboration and teamwork with learners~\cite{apoki2022,an2025impacts}. These so-called learning companions do not act as knowledge authorities, but as equal partners. 

Human-AI-collaboration or human-AI-teaming (HAIT) is defined as a synergistic partnership between humans and AI to achieve a common goal \cite{berretta2023}. Humans and AI should complement each other with their respective strengths~\cite{wang2026overloaded}. Humans contribute to intentionality, creativity, and ethical reasoning, while AI systems outperform humans in data processing, pattern recognition, and the ability to generate context-dependent information or suggestions~\cite{Dignum2017, engelhardt2025}. 

Collaboration has been shown to enhance learning by promoting active engagement, perspective-taking, and deeper cognitive processing~\cite{dillenbourg1999, johnson2009}. In collaborative learning environments, learners co-construct knowledge through discussion, observation, and negotiation of strategies, which facilitates reflection and the integration of multiple viewpoints. Observing and comparing one’s own actions with those of peers or, in the context of human-AI collaboration, with an adaptive AI partner, supports both explicit and implicit learning by providing opportunities for imitation, error correction, and iterative improvement~\cite{bandura1977, Chi2009}. Furthermore, collaboration can increase motivation and engagement, as learners' needs for relatedness are met when working within a social or interactive context that provides immediate feedback and shared responsibility for achieving goals~\cite{sawyer2005}.

\section{Educational design of our serious game}
\label{subsec:Game}
To investigate how different levels of AI-based scaffolding affect learning in serious games, we developed Qookies as a theory-based serious game designed to make abstract concepts of quantum technology accessible to learners without prior knowledge of mathematics or physics. Qookies targets a heterogeneous audience, including middle school and high school students, university students, and the general public. Consequently, it must introduce ideas of quantum technologies in an intuitive, visual, and experiential way. This objective poses specific demands on game design and instructional methods. %, but if done correctly, these foundations can be applied to many other topics as well.
To address this, learners progress through the game by solving interactive educational puzzles, as shown in ~\autoref{fig:ElementPuzzle}, which are complemented by short conceptual units (see ~\autoref{fig:ElementKnowledgeUnit}) that introduce underlying quantum technology concepts in simple and accessible language. To create effective learning experiences, we based the design of Qookies on several pedagogical foundations, which guided decisions on how to structure content, sequence puzzles, and implement adaptive AI support. Below, we explain how these principles from learning theories are integrated in the game design.
\begin{enumerate}
\item \textit{Personalization}: Personalization can be implemented in two different ways: adaptable or adaptive \cite{chernikova2025personalization}. In an \textit{adaptable} environment, the learner has multiple choices to personalize their learning path and sequence of content. In an \textit{adaptive} learning environment, the system personalizes the learning content based on certain information, such as learners’ characteristics or a prompt from the learner. Here, the scaffolding utilizes both types of personalization: learners can decide when to seek help and what help they would like (adaptable), while learning support is given based on the progress within a level and previously provided information (adaptive). In this way, learning support only occurs when the learner needs it, thereby avoiding unnecessary and redundant information.
\item \textit{Scaffolding}: Learning support is integrated through verbal (text-based) and visual (action-based) scaffolding. Verbal scaffolding is provided via a generative AI-based chatbot (Llama 3.1 70B) that is prompted in the background with the underlying concepts that are targeted in a level to avoid hallucinations; however, it does not know the solution for the level. In this way, it can explain concepts, clarify steps, and offer hints via chat, without revealing solutions. 

Visual scaffolding is realized  through an AI model based on one-shot reinforcement learning \cite{schrodt2025}. This model is embedded in an AI architecture that combines reinforcement learning and a large language model via an embodiment interface, grounding dialogue and action generation in the current game state. The reinforcement learning component is initialized without prior task-specific knowledge and learns from gameplay observation, enabling co-learning dynamics between the player and the AI-controlled NPC. In this way, the AI learns to perform actions that are likely to support solving the level and defers to the player when its knowledge is insufficient. By tightly coupling symbolic reasoning to grounded game-state representations and action plans, the architecture keeps dialogue and actions aligned with game mechanics and reduces the risk of hallucinations.

Both types of scaffolding are carried out through the NPC in the game. Figure~\ref{fig:ElementInteraction} illustrates these two forms of AI-based scaffolding
implemented through the NPC: verbal scaffolding via the chat interface
(\textit{``Talk to''}) and visual, action-based scaffolding via the
\textit{``Do something''} option, which allows the AI to perform in-game actions.
\item \textit{ICAP}: The game encourages \textit{active} engagement with the learning content in terms of the ICAP framework. In this way, it anchors understanding in specific experiences and naturally decomposes complex ideas into intuitive steps. For example, instead of telling learners what fluorescence means, the game allows them to work with the concept directly. An example of such active manipulation is shown in Figure~\ref{fig:ElementPuzzle}, where players adjust a qubit’s state on a Bloch sphere to solve a level. Moreover, when solving a level, learners need to apply their knowledge and construct a sequence of solution steps, which can be considered a form of \textit{constructive} engagement within the ICAP framework. Additionally, verbal scaffolding allows the communication and discussion about level-relevant concepts, potentially including turn-taking events. Therefore, it qualifies as an \textit{interactive} engagement with the learning content.
\item \textit{Extraneous Cognitive Load}: In the design of the game, it was avoided to include any kind of salient learning-irrelevant objects that might require the learner to evaluate the relevance of the object for solving the level, potentially leading to an increase in the extraneous cognitive load. This means that learning-irrelevant objects, such as a chair, elements on the desk or plants, do not exhibit a colorful contrast to the environment and they are not manipulative.
\item \textit{Serious Games}: From the learner’s perspective, conceptual and procedural understanding, as well as the application of knowledge, are not the explicit goal of the game. Instead, the perceived goal is to solve escape-room-based levels and progress through the game, which constitutes an artificial challenge and therefore qualifies the game as a serious game.
%\item \textit{Cognitive theory of multimedia learning}: In the game, we integrated {\em learning nuggets}, that is, small, self-contained learning units that explicitly relate the steps required to solve a level to the underlying conceptual and procedural knowledge. These learning nuggets contain explaining text as well as content-related visual representations, making use of the multimedia principle \cite{mayer2024past}. Moreover, concept-related objects are manipulative to support understanding of their functions; therefore, they stimulate active processing. 
\item \textit{Multiple External Representations (MERs)}: The learning content is encoded via two types of representations,
visual-graphical (e.g., the Bloch-Sphere) and verbal text-based representations, and both representations are manipulative and needed for solving a level. 
  
\item \textit{Collaborative Knowledge construction}: Both AI models, which are incorporated in the AI-based NPC character in the game, do not contain the solution of the levels at the beginning of each level. However, both AI models that control the NPC character learn and adapt their knowledge and action model via reinforcement learning based on the actions of the player and the NPC character. In this way, the NPC character and the player learn together and create new understandings and solutions, which is similar to collaborative knowledge construction among peers \cite{fischer2002fostering}. At the same time, we mitigate an overreliance of the learners on the generative AI output and a low cognitive activity of the learners, because the AI models do not know the solution of the level at the beginning of each level and must learn with the learner \cite{kosmyna2025your,krupp2024unreflected}. 

\end{enumerate}
	
Please note that in the current study, only aspects 1 to 3 are manipulated between conditions. All other design mechanisms are implemented identically across conditions and are expected to contribute equally to learning.

%\section{The game Qookies}
%\label{subsec:Game}
%The aim of the game is to provide an easy introduction to the topic of quantum physics and quantum technologies. It is designed to explain basic concepts of quantum physics and how they are used in quantum technologies. It is intended not only for pupils or students, but also for the general public. This objective poses several requirements for level design and content preparation. Due to the heterogeneous target group, it is not possible to assume a uniform level of knowledge. Therefore, the game should be designed in a way that requires no prior knowledge of mathematics or physics to understand the content. 

\begin{figure}[h]
    \centering
    \includegraphics[width=1.0\linewidth]{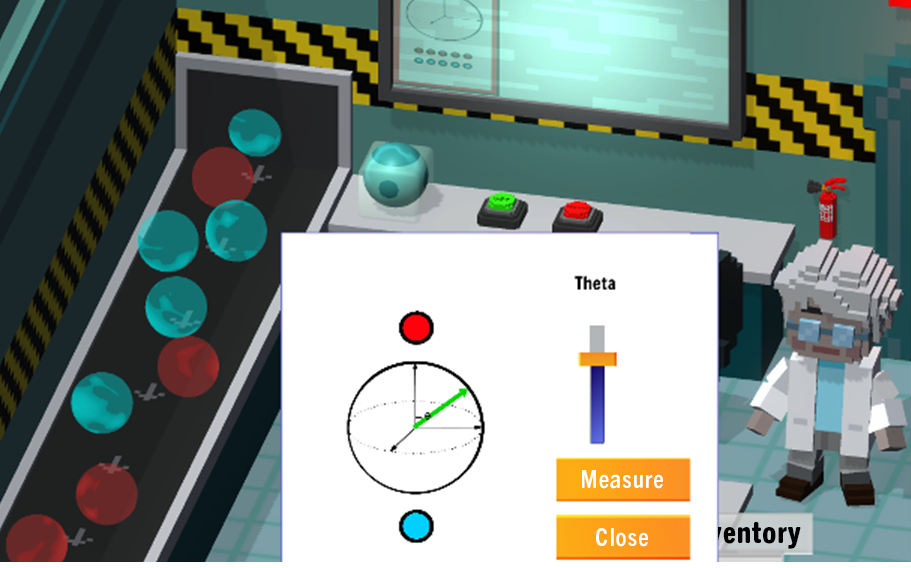}
    \caption{Example of an interactive educational puzzle in Qookies. Learners progress through the game by manipulating visual elements and solving level-specific challenges that embody core concepts of quantum technologies.}
    \label{fig:ElementPuzzle}
\end{figure}

%The composition of the game is based on three design elements. Firstly, we utilize the advantages of the medium by displaying the concepts of quantum mechanics as interactive tasks that need to be solved to progress in the game (cf.~\autoref{fig:ElementPuzzle}). 
%Secondly, short knowledge units provide the theoretical background to the levels in simple language~(cf.~\autoref{fig:ElementKnowledgeUnit}). 

\begin{figure}[h]
    \centering
    \includegraphics[width=1.0\linewidth]{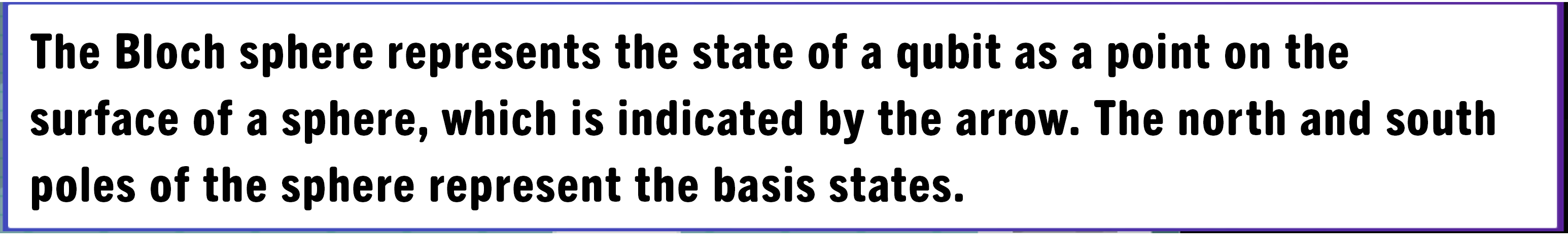}
    \caption{Brief learning unit providing concise background information on a quantum technology concept. These units introduce theoretical ideas in simple and accessible language and are directly connected to the interactive puzzles.}
    \label{fig:ElementKnowledgeUnit}
\end{figure}

%And thirdly, the players can interact with the AI-controlled non-player-character (NPC), which supports them in different ways. Depending on the conditions to which study participants are assigned, the NPC answers questions, offers tips for their further progress, or can be asked to demonstrate combinations of actions derived by the AI from the players' previous gameplay behavior~(cf.~\autoref{fig:ElementInteraction}). 

\begin{figure}[h]
    \centering
    \includegraphics[width=1.0\linewidth]{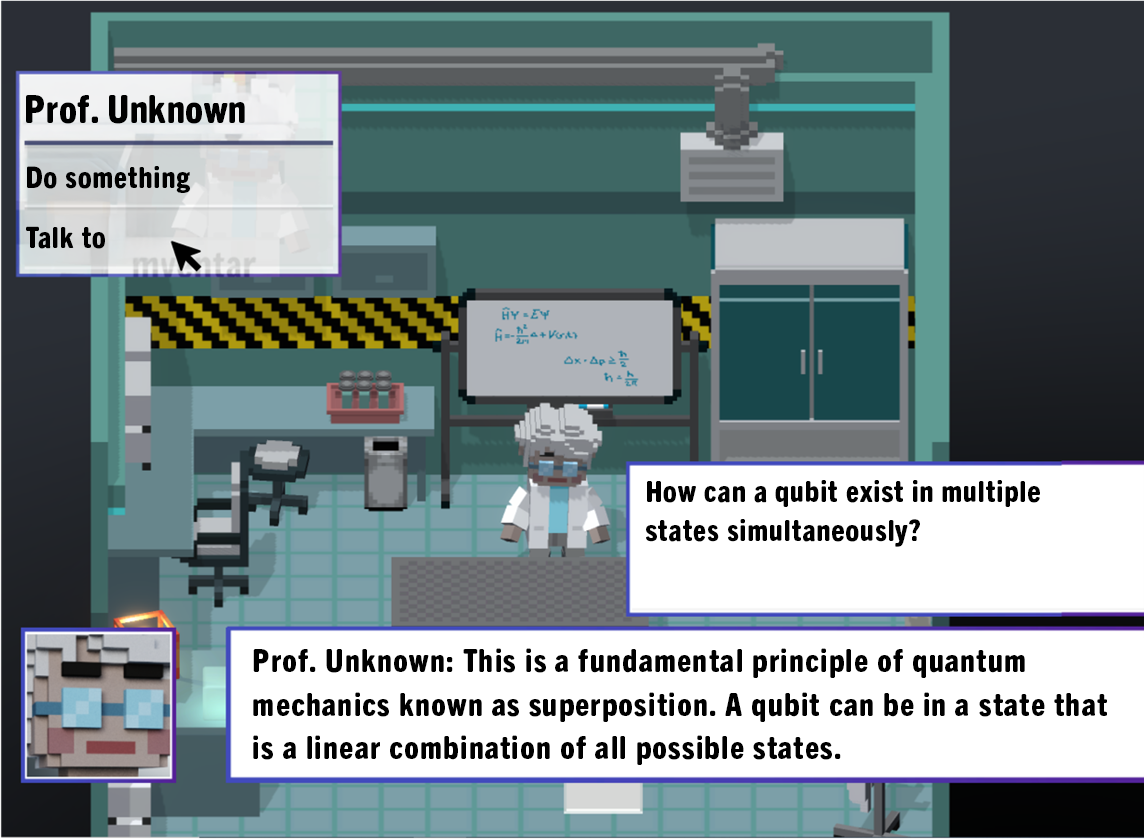}
    \caption{AI-based non-player character (NPC) providing learning support in Qookies. Learners can request verbal explanations via chat through the ``Talk to'' menu or trigger visual (action-based) scaffolding through the ``Do something'' menu, allowing the AI to demonstrate potentially helpful in-game actions.}
    \label{fig:ElementInteraction}
\end{figure}

\iffalse
\begin{figure}[h!]
    \centering
    % Linke Spalte: Zwei kleinere Bilder übereinander
    \begin{minipage}[b]{0.45\textwidth} % 45% der Textbreite für die linke Spalte
        \begin{subfigure}{\textwidth}
            \includegraphics[width=\textwidth]{figs/Element_Puzzle_formatted_englisch.png}
            \caption{Interactive puzzle about the Bloch-Sphere.}\label{fig:ElementPuzzle}
        \end{subfigure}
        \vspace{\fill} % Abstand zwischen den beiden kleinen Bildern
        \begin{subfigure}{\textwidth}
            \includegraphics[width=\textwidth]{figs/Element_KnowledgeUnit_englisch.png} 
            \caption{Textual knowledge unit.}\label{fig:ElementKnowledgeUnit}\vspace{0pt}
        \end{subfigure}
    \end{minipage}
    \hfill % Horizontaler Abstand zwischen den Spalten
    % Rechte Spalte: Großes Bild
    \begin{minipage}[b]{0.53\textwidth} % 55% der Textbreite für die rechte Spalte
        \begin{subfigure}{\textwidth}
            \includegraphics[width=\textwidth, height=5.67cm]{figs/Element_InteractionNPC_formatted_englisch.png}
            \caption{Chat interaction with the NPC.}\label{fig:ElementInteraction}\vspace{0cm}
        \end{subfigure}
    \end{minipage}

\caption{The three design elements used in the game.}
\end{figure}
\fi

\section{Research questions and hypotheses}
\label{sec:RQ}

The design of the game incorporates two forms of AI-mediated scaffolding, verbal and visual, as described in the Section~\ref{subsec:Game}. These forms differ in how instructional support is provided to the player and are intended to support learning while preserving opportunities for active problem solving. Therefore, the main research question of this study is:
%Game-based learning has been extensively researched; however, the implementation of AI as a support mechanism in serious games is still in its infancy and therefore remains largely unexplored in terms of its potential and practical applications. If a game is supplemented with an AI-controlled NPC, this NPC can support players in various ways through scaffolding. It remains to be investigated which type of scaffolding is particularly beneficial in this context. The main research question of this study is therefore:

\begin{itemize}[label={\textbf{RQ1}}, leftmargin=3em]
    \item How can an AI character effectively support a player in learning quantum technology content?
\end{itemize}

Based on scaffolding theory, we hypothesize that cognitive resources freed up by scaffolding can be leveraged by learners to gain a deeper understanding of the content. Therefore, the hypothesis regarding learning gains is:

\begin{itemize}[label={\textbf{H1}}, leftmargin=3em]
    \item Participants with access to a verbal and visual AI-based scaffolding have a greater learning gain from the serious game than participants with access to only an AI-based verbal scaffolding or without scaffolding.
\end{itemize}

The second research question relates to a mechanism underlying the learning process. Cognitive load refers to the amount of resources used in working memory due to the inherent difficulty of a task, the representation of that task, and the formation of new schemata. Scaffolding can reduce cognitive load and free up cognitive resources, allowing for a deeper understanding. Additional elements, such as interacting with the NPC, which are not directly necessary for the learning process, could, in turn, increase cognitive load and hinder the learning process. Therefore, the second research question is:

\begin{itemize}[label={\textbf{RQ2}}, leftmargin=3em]
    \item How does the AI character influence the player's cognitive load?
\end{itemize}

%The study compares three conditions. Two groups receive verbal (chat AI) or verbal and nonverbal (chat and action AI) support from the NPC.
The verbal instructions provided by the NPC's chat function can direct the players' attention and thus reduce the amount of information they need to process. Nevertheless, players must identify the objects described independently and perform the necessary actions to solve the task themselves. In comparison, the action AI can provide non-verbal cues by performing actions itself. This makes it immediately apparent to observers which object is relevant and how it must be used to solve the task. Based on this rationale, we expect that relative to the verbal only condition, the  verbal and visual scaffolding condition should reduce extraneous cognitive load.
%This reduces element interactivity and decreases the ECL in Condition~2 compared to Condition~1 and the control group. 
Therefore, the hypothesis regarding cognitive load is as follows: 

\begin{itemize}[label={\textbf{H2}}, leftmargin=3em]
    \item Participants with access to a verbal and visual scaffolding have a lower cognitive load when playing the serious game than participants with access to only a verbal scaffolding or without AI support.
\end{itemize}

%The cognitive resources freed up by scaffolding can be leveraged by learners to gain a deeper understanding of the content. Therefore, the hypothesis regarding learning gains is:

%\begin{itemize}[label={\textbf{H1}}, leftmargin=3em]
%    \item Participants with access to a chat-and action-AI (Condition~2) have a greater learning gain from the serious game than participants with access to only a chat-AI (Condition~1) or without AI support (control group).
%\end{itemize}

This study was pre-registered (\url{https://aspredicted.org/q2vw-9m72.pdf}).

\section{Method}
\label{sec:Method}
To establish the necessary sample size, we performed an a priori power analysis using G*Power 3.1~\cite{faul2007}. Because multiple tests were used for the analysis, the power was calculated for every test, and the highest number of participants necessary was chosen. The one-way fixed effects analysis of variance (ANOVA) was decisive. In order to achieve a power of $0.80$ in this analysis, a sample size of 160 is required. For the calculation, we used a medium effect size ($f = 0.25$) and the alpha error probability $\alpha = 0.05$.

The study was approved by the Institution's Ethics Committee and we obtained informed consent from the participants.

%The study was approved by the Ethics Committee of the Faculty of Mathematics, Computer Science, and Statistics at Ludwig-Maximilians-Universität in Munich, and we obtained informed consent from the participants.

\subsection{Participants}
\label{subsec:Participants}
After data collection, 152 complete data sets were available for analysis. Participants were initially assigned to one of three game designs: 47 to a control condition (no AI support), 50 to the verbal scaffolding condition (chat-based AI), and 55 to the combined verbal and visual scaffolding condition (with AI-based actions for the visual component). A detailed description of these scaffolding designs is provided in Section~\ref{subsec:Game}. However, after analyzing participants’ actual usage of the AI agent, the group assignments required adjustments.

Seventeen participants who were originally assigned to the verbal scaffolding group were reassigned to the control group because they did not engage with the AI using the ``Talk to'' function. Similarly, two participants originally assigned to the combined verbal and visual scaffolding group were reassigned to the verbal scaffolding group because they interacted with the AI through chatting but did not use the ``Do something'' function. Eleven participants from the combined verbal and visual scaffolding group neither used the ``Do something'' function nor engaged in chatting and were therefore reassigned to the control group.

A detailed overview of the initial and corrected group distributions, including gender and occupation breakdowns, is provided in table~\ref{tab:cohort}.

\begin{table*}[t]
\centering
\small
\setlength{\tabcolsep}{4pt}
\begin{tabular}{p{6.2cm}ccc}
\toprule
 & \textbf{Control} & \textbf{Verbal} & \textbf{Verbal+Visual} \\
\midrule
Participants (initial) & 47 & 50 & 55 \\
Participants (final)   & 75 & 35 & 42 \\
\addlinespace[0.35em]

Gender (m/f/d), initial & 26/20/1 & 24/23/3 & 31/20/4 \\
Gender (m/f/d), final   & 39/31/5 & 19/16/0 & 23/16/3 \\
\addlinespace[0.35em]

\makecell[l]{Occupation, initial\\(pupil/student/employee/other)}
 & 35/7/3/2 & 26/27/4/3 & 34/15/4/2 \\
\makecell[l]{Occupation, final\\(pupil/student/employee/other)}
 & 53/13/4/5 & 20/12/3/0 & 22/14/4/2 \\
\bottomrule
\end{tabular}
\caption{Participant distribution by group, including gender (m = male, f = female, d = diverse) and occupation. Values in the first row reflect the initial assignment at the start of the study; values in the second row reflect the final group distribution after reassignment based on actual AI usage.}
\label{tab:cohort}
\end{table*}

\subsection{Study design}
\label{subsec:StudyDesign}
The entire experiment was conducted on computer devices and divided into three phases. Participants first completed a pretest, then played the game for approximately 25 minutes, and finally answered a posttest. The experiment lasted approximately 50 minutes in total. 
The study used a one-factor between-groups design to compare different types of player-AI interactions. The independent variable was the amount of scaffolding offered by the AI. The dependent variables consisted of:
\begin{itemize}
    \item pre- and posttest responses regarding conceptual understanding; self-developed multiple choice test comprising 16 items, each with four possible answers, only one of which is correct (see~\ref{app:CU_questionnaire}).
    \item self-reported cognitive load in the post-test; assessed using nine out of ten items from~\cite{leppink2013}, which were adapted to the topic of serious games  (see~\ref{app:CL_questionnaire}). 
    The items were rated on a Likert-scale ranging from 0 to 6 (0 meaning \textit{not at all the case} and 6 meaning \textit{completely the case}).
\end{itemize}
Additionally, the participants' age, occupation, and gender were collected.

\subsection{Data analysis}
\label{subsec:DataAnalysis}
In the current study, repeated measures analysis of variance (ANOVA) was used to examine differences in concept understanding scores between pre- and post-tests across the control group and the two experimental conditions. A one-way ANOVA was performed to test for group differences in ICL, ECL, and GCL. If significant differences were found, a Bonferroni post-hoc test was used to test for differences between the groups or the measurement time points. The data analysis was performed using the R programming language.

To analyze the interaction with the AI, user queries via chat were categorized by two independent raters. The categorization comprises three main categories, each with its own subcategories. The first category distinguishes between queries that are level-related and those that are not. The former are questions and statements that refer to a specific level, objects contained therein, or situations that occur there. Non-level-related queries are general or overarching questions and statements that do not reference a specific level. The second category refers to the phrasing used by users and distinguishes between (follow-up) questions and statements or requests. (Follow-up) Questions are defined as inputs that are phrased as questions or have a clarifying, inquiring function. Statements or requests are understood to be all inputs that are phrased as statements, messages, or direct requests. The third category is based on the ICAP framework and distinguishes between the different levels of user engagement. 
The subcategory “underlying concepts” included queries in which participants asked about basic principles, mechanisms, theory, or definitions without formulating them in their own words. This corresponds to the lowest level of the ICAP framework (passive). If participants explicitly or implicitly asked for advice on how to proceed in a particular situation, this was assigned to the subcategory “recommended actions.” This is seen as an indicator of independent action, as required for level 2 (active) of the ICAP framework. The third (constructive) and fourth (interactive) levels of the framework were addressed in the subcategories “Hypothesis/idea formation” and “Discussion.” Statements that cannot be assigned to any of the subcategories described are summarized under “Other.” Differences in categorizations were solved in a discussion between the raters. A definition of the categories can be found in Table~\ref{tab:CategoriesAIInteractions}. 

\begin{table*}
\centering
\begin{tabularx}{\textwidth}{l p{3.9cm} >{\raggedright\arraybackslash}X}
\textbf{Category} & \textbf{Subcategory} & \textbf{Description} \\
\toprule
Reference & Level-related &
Questions and statements that refer to a specific level, objects contained therein, or situations that occur there. \\
& Non-level-related &
General or overarching questions and statements without reference to a specific level. \\
\addlinespace[0.6em]
\midrule
\addlinespace[0.6em]
Phrasing & (Follow-up) questions &
All entries that are formulated as questions or have a clarifying, inquiring function. \\
& Statements/requests &
All entries that are formulated as statements, notifications, or direct (explanatory) prompts. \\
\addlinespace[0.6em]
\midrule
\addlinespace[0.6em]
Engagement & Underlying concepts &
All entries that ask about basic principles, mechanisms, theories, or definitions without formulating them in your own words. \\
& Recommendations for~actions &
Explicit or implicit requests for specific advice, guidance, or suggestions on how to proceed in a particular situation. \\
& Hypothesis/idea formation &
Formulating your own assumptions, conjectures, or ideas. \\
& Discussion &
Inputs that refer to previous statements, arguments, or content; these are commented on, evaluated, expanded upon, or questioned. \\
& Other & \\
\bottomrule
\end{tabularx}
\caption{Categories for classifying user conversations with the AI character during gameplay.}
\label{tab:CategoriesAIInteractions}
\end{table*}

\section{Results}
\label{sec:Results}

\subsection{Conceptual Understanding of Quantum Technologies}

\begin{figure}[ht]
  \centering
  \resizebox{\columnwidth}{!}{
  \input{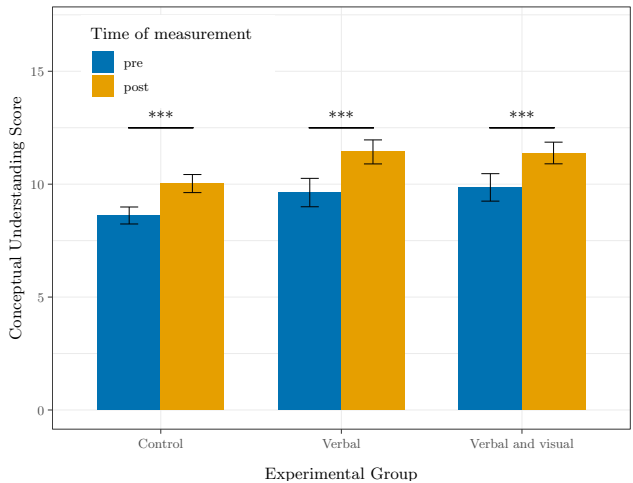}}
  \caption{Pre- and post-test scores in conceptual understanding (maximum score = 16) by group (control: no scaffolding; verbal scaffolding: chat-based; verbal and visual scaffolding: chat- and action-based). Asterisks indicate statistical significance: *** $p$ \textless 0.001, ** $p$ \textless 0.01, * $p$ \textless 0.05.}
  \label{fig:LearningGain_Barchart}
\end{figure}

Figure~\ref{fig:LearningGain_Barchart} shows the points achieved by the participants in the concept test depending on the group and measurement time.
The analysis did not reveal any significant difference in the pre-tests between the groups ($F(2,149)=1.98$, $p=0.141$%, $\eta^2=0.03$
). However, the repeated measures ANOVA showed a significant main effect of time (pre-test vs. post-test) on conceptual understanding ($F(1,149) = 58.85$, $p < .001$, $\eta^2=0.28$). Bonferroni post-hoc tests revealed a significant gain in points achieved from pre- to post-test for all groups: the control group ($t(149)=-4.968$, $p<.0001$, $d = 0.42$), the verbal scaffolding group ($t(149)=-4.323$, $p<.0001$, $d = 0.52$), and the verbal and visual scaffolding group ($t(149)=-4.009$, $p=.0001$, $d= 0.43$). The median increased from 9 to 11 points in the control group, from 10 to 12 points in the verbal scaffolding group, and from 11 to 13 points in the verbal and visual scaffolding group. 
There was no significant interaction effect between group and time ($F(2,149)=0.54$, $p=.583$%, $\eta^2=0.0039$
), indicating that the results of the post-tests in the three groups do not differ significantly from one another. 
An overview of the test results can be found in Table~\ref{tab:resultsLearning}.

\begin{table*}
    \centering
    \begin{tabularx}{\hsize}{X|c|c|c}
        Test & $F/t$ & $p$ & $\eta^2/d$\\
        \hline
        ANOVA & & & \\
        \hspace{10mm}one-way, pre-tests & $F(2,149)=1.98$ & 0.141 & %$\eta^2=0.03$
        \\
        \hspace{10mm}repeated measures, main effect: time & $F(1,149)=58.85$ & $<0.001$*** & $\eta^2=0.28$\\
        \hspace{10mm}repeated measures, interaction effect: group x time & $F(2,149)=0.54$ & $0.583$ & %$\eta^2=0.004$
        \\
        Post-hoc Tests (Bonferroni), pre vs. post &  &  & \\
        \hspace{10mm}Control group (no AI) & $t(149)=-4.968$ & $<0.001$*** & $d=0.42$\\
        \hspace{10mm}Verbal scaffolding (chat AI) & $t(149)=-4.323$ & $<0.001$*** & $d=0.52$\\
        \hspace{10mm}Verbal sand visual scaffolding (chat and action AI) & $t(149)=-4.009$ & $<0.001$*** & $d=0.43$\\
    \end{tabularx}
    \caption{An overview of the results of the conceptual understanding analysis. For the repeated measures ANOVAs with main effects and interactions, the one-way ANOVA, as well as Bonferroni-corrected post-hoc tests, are specified, along with the test statistics, $p$-values, and effect sizes.}
    \label{tab:resultsLearning}
    \end{table*}

    \begin{table*}
    \centering
    \begin{tabularx}{\hsize}{X|c|c|c}
        Test & $F/t$ & $p$ & $\eta^2/d$\\
        \hline
        ANOVA & & & \\
        \hspace{10mm}one-way, ICL & $F(2,148)=4$ & 0.02* & $\eta^2=0.051$\\
        \hspace{10mm}one-way, ECL & $F(2,148)=0.038$ & $0.96$ & %$\eta^2=0.0005$
        \\
        \hspace{10mm}one-way, GCL & $F(2,147)=0.87$ & $0.42$ & %$\eta^2=0.012$
        \\
        Post-hoc Tests (Bonferroni), ICL group comparison &  &  & \\
        \hspace{10mm}Control group - Verbal scaffolding & $t(148)=-0.867$ & $1$ & %$d=0.18$
        \\
        \hspace{10mm}Control group - Verbal and visual scaffolding & $t(148)=2.235$ & $0.081$ & %$d=0.43$
        \\
        \hspace{10mm}Verbal scaffolding - Verbal and visual scaffolding  & $t(148)=2.664$ & 0.026* & $d=0.60$\\
    \end{tabularx}
    \caption{An overview of the results of the cognitive load analysis. For the one-way ANOVAs as well as Bonferroni-corrected post-hoc tests, the test statistics, $p$-values, and effect sizes are specified.}
    \label{tab:resultsCL}
    \end{table*}

\subsection{Cognitive Load}
\label{subsec:ResultsRQ2}
\begin{figure}[ht]
  \centering
  \resizebox{\columnwidth}{!}{%
  % Created by tikzDevice version 0.12.6 on 2025-11-12 13:55:26
% !TEX encoding = UTF-8 Unicode
\begin{tikzpicture}[x=1pt,y=1pt]
\definecolor{fillColor}{RGB}{255,255,255}
\path[use as bounding box,fill=fillColor,fill opacity=0.00] (0,0) rectangle (383.03,289.08);
\begin{scope}
\path[clip] (  0.00,  0.00) rectangle (383.03,289.08);
\definecolor{drawColor}{RGB}{255,255,255}
\definecolor{fillColor}{RGB}{255,255,255}

\path[draw=drawColor,line width= 0.6pt,line join=round,line cap=round,fill=fillColor] (  0.00,  0.00) rectangle (383.03,289.08);
\end{scope}
\begin{scope}
\path[clip] ( 27.31, 30.69) rectangle (377.53,283.58);
\definecolor{fillColor}{RGB}{255,255,255}

\path[fill=fillColor] ( 27.31, 30.69) rectangle (377.53,283.58);
\definecolor{drawColor}{gray}{0.92}

\path[draw=drawColor,line width= 0.3pt,line join=round] ( 27.31, 83.98) --
	(377.53, 83.98);

\path[draw=drawColor,line width= 0.3pt,line join=round] ( 27.31,167.58) --
	(377.53,167.58);

\path[draw=drawColor,line width= 0.3pt,line join=round] ( 27.31,251.18) --
	(377.53,251.18);

\path[draw=drawColor,line width= 0.6pt,line join=round] ( 27.31, 42.18) --
	(377.53, 42.18);

\path[draw=drawColor,line width= 0.6pt,line join=round] ( 27.31,125.78) --
	(377.53,125.78);

\path[draw=drawColor,line width= 0.6pt,line join=round] ( 27.31,209.38) --
	(377.53,209.38);

\path[draw=drawColor,line width= 0.6pt,line join=round] ( 92.98, 30.69) --
	( 92.98,283.58);

\path[draw=drawColor,line width= 0.6pt,line join=round] (202.42, 30.69) --
	(202.42,283.58);

\path[draw=drawColor,line width= 0.6pt,line join=round] (311.87, 30.69) --
	(311.87,283.58);
\definecolor{fillColor}{RGB}{0,114,178}

\path[fill=fillColor] ( 54.67, 42.18) rectangle ( 80.21,182.55);

\path[fill=fillColor] (164.12, 42.18) rectangle (189.65,118.25);

\path[fill=fillColor] (273.56, 42.18) rectangle (299.10,170.45);
\definecolor{fillColor}{RGB}{230,159,0}

\path[fill=fillColor] ( 80.21, 42.18) rectangle (105.75,194.45);

\path[fill=fillColor] (189.65, 42.18) rectangle (215.19,118.62);

\path[fill=fillColor] (299.10, 42.18) rectangle (324.63,155.94);
\definecolor{fillColor}{RGB}{204,121,167}

\path[fill=fillColor] (105.75, 42.18) rectangle (131.28,153.65);

\path[fill=fillColor] (215.19, 42.18) rectangle (240.73,115.83);

\path[fill=fillColor] (324.63, 42.18) rectangle (350.17,158.87);
\definecolor{drawColor}{RGB}{0,0,0}

\path[draw=drawColor,line width= 0.6pt,line join=round] ( 63.79,190.16) --
	( 71.09,190.16);

\path[draw=drawColor,line width= 0.6pt,line join=round] ( 67.44,190.16) --
	( 67.44,174.95);

\path[draw=drawColor,line width= 0.6pt,line join=round] ( 63.79,174.95) --
	( 71.09,174.95);

\path[draw=drawColor,line width= 0.6pt,line join=round] (173.24,124.13) --
	(180.53,124.13);

\path[draw=drawColor,line width= 0.6pt,line join=round] (176.89,124.13) --
	(176.89,112.37);

\path[draw=drawColor,line width= 0.6pt,line join=round] (173.24,112.37) --
	(180.53,112.37);

\path[draw=drawColor,line width= 0.6pt,line join=round] (282.68,177.58) --
	(289.98,177.58);

\path[draw=drawColor,line width= 0.6pt,line join=round] (286.33,177.58) --
	(286.33,163.31);

\path[draw=drawColor,line width= 0.6pt,line join=round] (282.68,163.31) --
	(289.98,163.31);

\path[draw=drawColor,line width= 0.6pt,line join=round] ( 89.33,205.84) --
	( 96.63,205.84);

\path[draw=drawColor,line width= 0.6pt,line join=round] ( 92.98,205.84) --
	( 92.98,183.07);

\path[draw=drawColor,line width= 0.6pt,line join=round] ( 89.33,183.07) --
	( 96.63,183.07);

\path[draw=drawColor,line width= 0.6pt,line join=round] (198.77,126.12) --
	(206.07,126.12);

\path[draw=drawColor,line width= 0.6pt,line join=round] (202.42,126.12) --
	(202.42,111.12);

\path[draw=drawColor,line width= 0.6pt,line join=round] (198.77,111.12) --
	(206.07,111.12);

\path[draw=drawColor,line width= 0.6pt,line join=round] (308.22,165.96) --
	(315.51,165.96);

\path[draw=drawColor,line width= 0.6pt,line join=round] (311.87,165.96) --
	(311.87,145.92);

\path[draw=drawColor,line width= 0.6pt,line join=round] (308.22,145.92) --
	(315.51,145.92);

\path[draw=drawColor,line width= 0.6pt,line join=round] (114.87,164.33) --
	(122.16,164.33);

\path[draw=drawColor,line width= 0.6pt,line join=round] (118.52,164.33) --
	(118.52,142.97);

\path[draw=drawColor,line width= 0.6pt,line join=round] (114.87,142.97) --
	(122.16,142.97);

\path[draw=drawColor,line width= 0.6pt,line join=round] (224.31,124.51) --
	(231.61,124.51);

\path[draw=drawColor,line width= 0.6pt,line join=round] (227.96,124.51) --
	(227.96,107.15);

\path[draw=drawColor,line width= 0.6pt,line join=round] (224.31,107.15) --
	(231.61,107.15);

\path[draw=drawColor,line width= 0.6pt,line join=round] (333.75,168.32) --
	(341.05,168.32);

\path[draw=drawColor,line width= 0.6pt,line join=round] (337.40,168.32) --
	(337.40,149.43);

\path[draw=drawColor,line width= 0.6pt,line join=round] (333.75,149.43) --
	(341.05,149.43);

\path[draw=drawColor,line width= 0.3pt,line join=round] ( 92.98,217.74) -- (119.25,217.74);

\path[draw=drawColor,line width= 0.3pt,line join=round] ( 92.98,217.74) -- (119.25,217.74);

\path[draw=drawColor,line width= 0.3pt,line join=round] ( 92.98,217.74) -- (119.25,217.74);

\path[draw=drawColor,line width= 0.3pt,line join=round] ( 92.98,217.74) -- (119.25,217.74);

\path[draw=drawColor,line width= 0.3pt,line join=round] ( 92.98,217.74) -- (119.25,217.74);

\path[draw=drawColor,line width= 0.3pt,line join=round] ( 92.98,217.74) -- (119.25,217.74);

\path[draw=drawColor,line width= 0.3pt,line join=round] ( 92.98,217.74) -- (119.25,217.74);

\path[draw=drawColor,line width= 0.3pt,line join=round] ( 92.98,217.74) -- (119.25,217.74);

\path[draw=drawColor,line width= 0.3pt,line join=round] ( 92.98,217.74) -- (119.25,217.74);

\node[text=drawColor,anchor=base,inner sep=0pt, outer sep=0pt, scale=  1.10] at (106.11,218.14) {*};
\definecolor{drawColor}{gray}{0.20}

\path[draw=drawColor,line width= 0.6pt,line join=round,line cap=round] ( 27.31, 30.69) rectangle (377.53,283.58);
\end{scope}
\begin{scope}
\path[clip] (  0.00,  0.00) rectangle (383.03,289.08);
\definecolor{drawColor}{gray}{0.30}

\node[text=drawColor,anchor=base east,inner sep=0pt, outer sep=0pt, scale=  0.88] at ( 22.36, 39.15) {0};

\node[text=drawColor,anchor=base east,inner sep=0pt, outer sep=0pt, scale=  0.88] at ( 22.36,122.75) {2};

\node[text=drawColor,anchor=base east,inner sep=0pt, outer sep=0pt, scale=  0.88] at ( 22.36,206.35) {4};
\end{scope}
\begin{scope}
\path[clip] (  0.00,  0.00) rectangle (383.03,289.08);
\definecolor{drawColor}{gray}{0.20}

\path[draw=drawColor,line width= 0.6pt,line join=round] ( 24.56, 42.18) --
	( 27.31, 42.18);

\path[draw=drawColor,line width= 0.6pt,line join=round] ( 24.56,125.78) --
	( 27.31,125.78);

\path[draw=drawColor,line width= 0.6pt,line join=round] ( 24.56,209.38) --
	( 27.31,209.38);
\end{scope}
\begin{scope}
\path[clip] (  0.00,  0.00) rectangle (383.03,289.08);
\definecolor{drawColor}{gray}{0.20}

\path[draw=drawColor,line width= 0.6pt,line join=round] ( 92.98, 27.94) --
	( 92.98, 30.69);

\path[draw=drawColor,line width= 0.6pt,line join=round] (202.42, 27.94) --
	(202.42, 30.69);

\path[draw=drawColor,line width= 0.6pt,line join=round] (311.87, 27.94) --
	(311.87, 30.69);
\end{scope}
\begin{scope}
\path[clip] (  0.00,  0.00) rectangle (383.03,289.08);
\definecolor{drawColor}{gray}{0.30}

\node[text=drawColor,anchor=base,inner sep=0pt, outer sep=0pt, scale=  0.88] at ( 92.98, 19.68) {ICL};

\node[text=drawColor,anchor=base,inner sep=0pt, outer sep=0pt, scale=  0.88] at (202.42, 19.68) {ECL};

\node[text=drawColor,anchor=base,inner sep=0pt, outer sep=0pt, scale=  0.88] at (311.87, 19.68) {GCL};
\end{scope}
\begin{scope}
\path[clip] (  0.00,  0.00) rectangle (383.03,289.08);
\definecolor{drawColor}{RGB}{0,0,0}

\node[text=drawColor,rotate= 90.00,anchor=base,inner sep=0pt, outer sep=0pt, scale=  1.10] at ( 13.08,157.13) {Cognitive Load Score};
\end{scope}
\begin{scope}
\path[clip] (  0.00,  0.00) rectangle (383.03,289.08);
\definecolor{fillColor}{RGB}{255,255,255}

\path[fill=fillColor] (237.44,208.95) rectangle (347.40,278.52);
\end{scope}
\begin{scope}
\path[clip] (  0.00,  0.00) rectangle (383.03,289.08);
\definecolor{drawColor}{RGB}{0,0,0}

\node[text=drawColor,anchor=base west,inner sep=0pt, outer sep=0pt, scale=  1.10] at (242.94,264.38) {Experimental Group};
\end{scope}
\begin{scope}
\path[clip] (  0.00,  0.00) rectangle (383.03,289.08);
\definecolor{fillColor}{RGB}{255,255,255}

\path[fill=fillColor] (242.94,243.35) rectangle (257.40,257.81);
\definecolor{fillColor}{RGB}{0,114,178}

\path[fill=fillColor] (243.66,244.07) rectangle (256.69,257.10);
\end{scope}
\begin{scope}
\path[clip] (  0.00,  0.00) rectangle (383.03,289.08);
\definecolor{fillColor}{RGB}{255,255,255}

\path[fill=fillColor] (242.94,228.90) rectangle (257.40,243.35);
\definecolor{fillColor}{RGB}{230,159,0}

\path[fill=fillColor] (243.66,229.61) rectangle (256.69,242.64);
\end{scope}
\begin{scope}
\path[clip] (  0.00,  0.00) rectangle (383.03,289.08);
\definecolor{fillColor}{RGB}{255,255,255}

\path[fill=fillColor] (242.94,214.45) rectangle (257.40,228.90);
\definecolor{fillColor}{RGB}{204,121,167}

\path[fill=fillColor] (243.66,215.16) rectangle (256.69,228.19);
\end{scope}
\begin{scope}
\path[clip] (  0.00,  0.00) rectangle (383.03,289.08);
\definecolor{drawColor}{RGB}{0,0,0}

\node[text=drawColor,anchor=base west,inner sep=0pt, outer sep=0pt, scale=  0.88] at (262.90,247.55) {Control};
\end{scope}
\begin{scope}
\path[clip] (  0.00,  0.00) rectangle (383.03,289.08);
\definecolor{drawColor}{RGB}{0,0,0}

\node[text=drawColor,anchor=base west,inner sep=0pt, outer sep=0pt, scale=  0.88] at (262.90,233.10) {Verbal};
\end{scope}
\begin{scope}
\path[clip] (  0.00,  0.00) rectangle (383.03,289.08);
\definecolor{drawColor}{RGB}{0,0,0}

\node[text=drawColor,anchor=base west,inner sep=0pt, outer sep=0pt, scale=  0.88] at (262.90,218.64) {Verbal and visual};
\end{scope}
\end{tikzpicture} }
  \caption{Self-reported intrinsic, extraneous, and germane cognitive load by group. Cognitive load was assessed using items adapted from \cite{leppink2013} (see~\ref{app:CL_questionnaire}) on a 0–6 Likert scale (low to high cognitive load). Groups differed in scaffolding: control (no AI support), verbal scaffolding (chat-based AI), and verbal and visual scaffolding (chat- and action-based AI). Asterisks indicate statistical significance: *** $p$ \textless 0.001, ** $p$ \textless 0.01, * $p$ \textless 0.05.}
  \label{fig:CL_Barchart}
\end{figure}
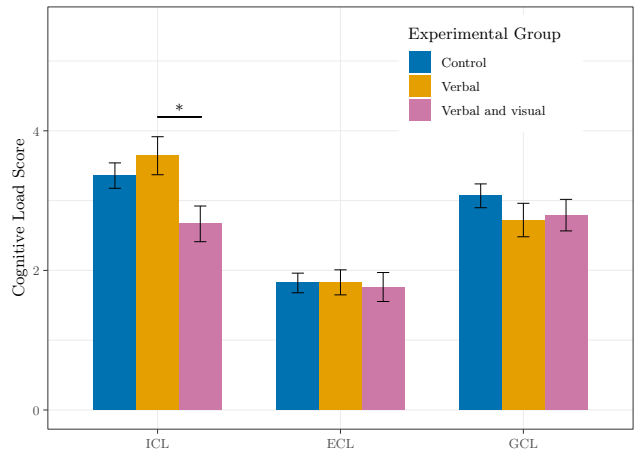

Looking at figure~\ref{fig:CL_Barchart}, the analysis showed no significant difference between the groups in ECL ($F(2,148)=0.038$, $p=0.962$%, $\eta^2=0.0005$
) and GCL ($F(2,147)= 0.87$, $p=0.421$%, $\eta^2=0.012$
). The ANOVA revealed a significant difference between the groups in ICL ($F(2,148)=4$, $p= 0.0203$, $\eta^2=0.051$). The Bonferroni post-hoc test showed no significant difference between the control group and the verbal scaffolding group ($t(148)=-0.87$, $p=1.00$%, $d=0.18$
) as well as the control group and the verbal and visual scaffolding group ($t(2,148)=2.235$, $p=.081$%, $d = 0.43$
), but a significant difference between the verbal scaffolding group and the verbal and visual scaffolding group ($t(148)=2.66$, $p=.026$, $d=0.60$). The results are displayed in figure~\ref{fig:CL_Barchart} and table~\ref{tab:resultsCL}.

\subsection{Interaction with AI}
\label{subsec:}
A total of 71 participants submitted 246 queries to the AI. The raters assigned each query to a subcategory from all main categories. The inter-rater reliability was $\kappa_1=0.79$ for Reference, $\kappa_2=0.78$ for Phrasing, and $\kappa_3=0.80$ for Engagement. According to~\cite{landis1977}, this is considered substantial agreement. 

In most cases, participants' queries were level-related and thus referred to a specific situation in the game, rather than asking general questions or making general statements. In addition, most queries were formulated as questions. The ratings in category 3 (Engagement) clearly indicate that these questions were primarily used to solicit advice for further action. A question that was categorized as level-related, (follow-up) question, and recommendation for action is ``\textit{How can you recognize a diamond with quantum properties?}”. ``\textit{Hi Prof. Unknown. I would like to know more about qubits. Please explain it starting from basics}'' on the other hand, is a non-level-related request to explain an underlying concept. 

No query to the AI was categorized as hypothesis or idea formation. Users therefore never formulated their own assumptions, conjectures, or ideas to the AI. 

Twelve queries were assigned to the discussion category. These include, for example, ``\textit{I don't have the laser in my inventory}''. %Even though this is clearly a response to a statement made by the AI, it does not meet the level of engagement required by the ICAP framework. For behavior to be considered interactive, it must at least be a constructive contribution, which is not the case here. The category formulated by the authors does not cover the highest level of engagement in the ICAP framework as intended.

Queries that were categorized as ``other'' are typically statements or questions like ``\textit{Hi}'' or ``\textit{How are you?}''. 

All categories and the number of queries are listed in table~\ref{tab:resultsNumberQueries}.

\begin{table}[ht]
\centering
\resizebox{\columnwidth}{!}{%
\begin{tabular}{llr}
\toprule
\textbf{Category} & \textbf{Subcategory} & \textbf{Number of queries} \\
\midrule
Reference 
    & Level-related          & 157 \\
    & Non-level-related      & 89  \\
\addlinespace[0.5em]
Phrasing 
    & (Follow-up) questions  & 175 \\
    & Statements/requests    & 71  \\
\addlinespace[0.5em]
Engagement 
    & Underlying concepts          & 20  \\
    & Recommendations for actions  & 124 \\
    & Hypothesis/idea formation    & 0   \\
    & Discussion                   & 12  \\
    & Other                        & 90  \\
\bottomrule
\end{tabular}%
}
\caption{Overview of the player--AI conversation categories and the frequency with which they occurred.}
\label{tab:resultsNumberQueries}
\end{table}

\section{Discussion}
\label{sec:Discussion}
\subsection{Learning about Quantum Technologies}
The analysis of the conceptual understanding in the pre- and post-tests shows that participants from all groups learned through the game. These results confirm the effectiveness of this serious game for learning quantum physics content, and they are consistent with findings from other studies~\cite{anupam2017, escanez2024}. Although there were differences between the groups, both in terms of prior knowledge and conceptual understanding after playing the game, these differences were not significant. %To confirm or refute our hypothesis that participants in {\color{red}the Chat-and-Action AI group} learn more, more participants would be needed. 

\subsection{Cognitive Load}
In the case of ECL and GCL, all three groups showed similar distributions. This suggests that each condition poses similar cognitive demands in terms of integrating new knowledge with prior knowledge and that none of the conditions introduced distracting features. Specifically, interaction with the AI was easy and did not require additional learning-irrelevant mental resources compared to participants who did not have access to these features. 

In comparison, participants from the verbal scaffolding group reported significantly higher ICL than participants from the verbal and visual scaffolding group. This indicates that verbal scaffolding alone via generative AI does not support learners in managing cognitive load. Only the combination of verbal and visual scaffolding was able to significantly reduce intrinsic cognitive load for players.

The initial hypothesis that participants in the verbal and visual scaffolding group would have less ICL than participants in the verbal scaffolding group or the control group was only partially confirmed. 
Participants who received the verbal and visual scaffolding report lower cognitive load in comparison to the verbal scaffolding group, but not in comparison to the control group. The latter finding is in agreement with prior observations that textual scaffolds in a VR environment do increase students' cognitive load. The former finding is in line with the multimedia principle \cite{mayer2002multimedia} and with recent frameworks on human-AI collaboration \cite{wang2026overloaded}. It is likely that participants in the verbal scaffolding group found it more challenging to interact with the AI via chat only. Reflecting on their own problems and formulating questions based on them could be a hurdle. In contrast, participants in the verbal and visual scaffolding group had the opportunity to alternatively receive visual input from the AI in these situations. The main difference here is that the help is not text-based, but instead an action is carried out by the AI character. Thus, visual scaffolding does not require learners to translate a partially abstract communication with the AI character into sequence of actions that lead to the solution
.

In contrast to the results in this work, the study presented by~\cite{wang2025} reports that school students using an LLM in a contextual game experienced significantly lower cognitive load than school students in a control group. \cite{wang2025} assesses cognitive load using questionnaires based on~\cite{paas1992}, who does not distinguish between ICL, ECL, and GCL, but rather between mental load and mental effort. In contrast, the AI models implemented in this study aimed to mitigate cognitive offloading of learning-relevant tasks to the LLM, reduced cognitive activity, and an unreflected acceptance of the output of the LLM, as observed earlier~\cite{krupp2024unreflected,kosmyna2025your}, by not knowing the correct solutions for the levels at the beginning of each level. Instead, the AI models learn alongside the learner. Such factors may also lead to a reduced cognitive load and low retention, but they do not seem to occur in the way the AI models were implemented in this work. 

Moreover, recent studies report that metacognitive scaffolding through generative AI in an educational game can increase GCL, suggesting a deeper construction of mental schemata \cite{ngu2025generative}. In contrast, our findings indicate that verbal cognitive scaffolding via generative AI alone did not significantly affect GCL, whereas the addition of visual AI scaffolding reduced ICL. Together, these findings suggest that different AI scaffolding implementations may influence different components of cognitive load.

\subsection{Interaction with AI}
The evaluation of the assignment to the groups, i.e., whether a learner who was in the verbal scaffolding group actually used the NPC's chat function, showed that only a fraction of the participants used the LLM as intended. To progress in the game, it was not necessary to use the AI character. %And even though the functionalities were explained in text, it is possible that this was overlooked. 
Another explanation could be that this use of AI in games is not widespread enough for its use to be intuitive for players. They may not have realized that once the features were introduced, they could get this support in every level. Participants also mentioned that they wanted to complete the levels on their own and that using AI would feel like cheating. This suggests that they may have perceived the levels as a desirable difficulty, which is considered valuable for cognition and learning \cite{bjork2020desirable}. Especially, in context of recent findings that students offload learning-relevant tasks and low cognitive activity when using AI in education \cite{krupp2024unreflected,stadler2024cognitive,kosmyna2025your}, the finding that students willingly invest mental effort in game-based learning environments instead of off-loading the tasks to AI is relevant for the current discussion on how to integrate AI into education. 

Another obstacle to using AI arose in the verbal and visual scaffolding group. Since the action AI for visual scaffolding first has to learn from the player, it cannot help in new situations. If this happens when players are trying out the AI for the first time, it may lead to the assumption that the AI is not helpful at all and may discourage them from using it again, even though it was described in this way to the learners. 

The evaluation of the categorization of player-AI conversations suggests that learners often interacted with the AI in ways consistent with a tutor-like role, primarily seeking answers or guidance when they were stuck. Players expect the AI to know the solution to the level, rather than perceiving it as a peer with different, but equally incomplete, knowledge. From this position, there was an expectation of infallibility on the part of the AI that it could not live up to. Similarly, learners may have gotten confused or frustrated by answers that were incorrect or impossible to implement.

%This may also be a reason why no joint idea formation takes place. 
In addition, the low frequency of interactions with the AI model might also be influenced by the game design, which may not have offered enough incentives or added value for using the AI model. 

%\subsection{Implications}
%\label{subsec:Implications}
%The results indicate that the NPC should be revised in several respects. First, the character design should be adjusted in order to move the AI away from the traditional tutor role and present it as a peer. To do this, the NPC needs a design that children and young people can identify with. Additionally, the chat function must be simplified to make it more accessible. To enable continuous exchange, the chat should not close after each request or response and have to be reopened in order to respond or ask a follow-up question.  A messenger chat format, which children and young people using smartphones are already familiar with, would be ideal for this purpose. 

%Second, it is important to introduce the individual AI functionalities step by step and to ensure that the level can only be successfully completed if these are used. “Reminder levels” should be implemented that require the use of the different functions and thus challenge players to use all of them. 

%Third, an additional function would be useful if the action model has not yet learned the solution from the player and therefore cannot act on its own. At this point, a function could be implemented that verbalizes the next steps without the player having to formulate their own question.  

\section{Conclusion}
\label{sec:Conclusion}

In this work, we studied two ways to implement AI scaffolding to support learning within a serious game. The results show that the game-based learning environment effectively supports the learning of concepts related to quantum technologies, regardless of the type of AI-based scaffolding. Par\-ti\-ci\-pants in all groups demonstrated significant learning gains, indicating that learning was robust across conditions, and no significant differences between the AI-supported conditions and the control group were observed. 

%We found that combining  verbal and visual scaffolding does not lead to higher learning outcomes, but does reduce intrinsic cognitive load. This suggests that visual scaffolding helps learners better manage cognitive load than verbal scaffolding alone.

The AI support was implemented such that the AI learned together with the learner based on the learner’s actions. This design contrasts with previous work in which students received AI support via free-form chatbots. In these cases, the authors observed that such approaches can lead to unreflected acceptance of AI-generated output by learners and reduced cognitive activity compared to a control condition without AI support~\cite{krupp2024unreflected,kosmyna2025your}. In contrast, our results indicate that because the AI companion learned alongside the player and could not provide immediate solutions, visual, action-based support reduced intrinsic cognitive load without encouraging cognitive offloading, even though it did not enhance content acquisition more than the other conditions. The verbal statements of the participants also suggest that the tasks in the serious game were perceived as a desirable difficulty. This suggests that AI companions designed as co-learners can support allocation of cognitive resources between learners and AI while preserving learners’ responsibility for conceptual understanding.

%collaborative learning between AI and the learners was an effective integration to avoid negative effects, such as unreflected acceptance of the output of the AI models, a cognitive offloading of learning-relevant subtasks, and a low cognitive activity, which might have challenged the learning outcomes in the AI-supported groups. 
%although adding visual to verbal scaffolding did not increase learning outcomes, it significantly reduced intrinsic cognitive load. This suggests that visual, action-based scaffolding supports learners primarily by regulating cognitive demands rather than by directly enhancing content acquisition. Because the AI companion could not provide immediate solutions and instead learned alongside the player, action-based support may have reduced intrinsic cognitive load without encouraging reliance on the AI. Taken together, these results indicate that AI companions designed as co-learners can support cognitive regulation while preserving learners’ responsibility for conceptual understanding.

Furthermore, the analysis of learner interactions with the AI character showed that the AI was primarily used for level-related questions and recommendations for action, with occasional deeper discussions. This highlights the value and importance of careful AI integration into serious game design. To fully exploit the potential of AI in serious games, character design, the guidance of chat-based interactions, and the integration of AI-supported actions within the gameplay need to be carefully aligned. Future research should therefore focus on generalizing the effects of the role of AI as a co-learner on learning outcomes and cognitive load found in this work to other settings and ensuring cognitively activating use. 

\begin{acknowledgments}
This work was supported by the Federal Ministry of Research, Technology and Space (BMFTR) under grant number [FKZ: 13N16725]. The authors are responsible for the content of this publication.
\end{acknowledgments}

\appendix
\section{Conceptual understanding questionnaire}
\label{app:CU_questionnaire}

This questionnaire is a self-developed test to assess conceptual understanding in the serious game Qookies. %It was originally developed and used in German and can be viewed in the supplementary material. For this paper, the questionnaire was translated into English. 
The multiple-choice questions have four options, only one of which is correct. The correct answer is marked in bold or, in the case of image answers, with asterisks. \\

\noindent Which statement about the energy of light is correct?
\begin{enumerate}[label=\alph*)]
\setlength{\itemsep}{0pt}
    \item Light with a long wavelength has higher energy than light with a short wavelength.
    \item \textbf{Blue light has higher energy than red light.}
    \item All colors in the visible spectrum have the same energy. 
    \item Ultraviolet light has a lower energy than infrared light.
\end{enumerate}

\noindent How can you determine that the emitted energy is different from that of the laser used for illumination?
\begin{enumerate}[label=\alph*)] \setlength{\itemsep}{0pt}
    \item By the change of temperature in the material.
    \item By the intensity of the emitted light.
    \item \textbf{By the different color of the fluorescent light compared to the laser.}
    \item By the duration of the fluorescence.
\end{enumerate}

\noindent What is the relationship between the energy of the emitted light and the energy of the absorbed light in a fluorescent material?
\begin{enumerate}[label=\alph*)] \setlength{\itemsep}{0pt}
    \item The energy of the emitted light is higher than that of the absorbed light.
    \item The energy of the emitted light is equal to the energy of the absorbed light.
    \item \textbf{The energy of the emitted light is lower than that of the absorbed light.}
    \item The energy of the emitted light and the energy of the absorbed light are unrelated.
\end{enumerate}

\noindent How do bits and Qubits differ?
\begin{enumerate}[label=\alph*)] \setlength{\itemsep}{0pt}
    \item Bits can only take the values 0 and 1, while Qubits can only take the value 1.
    \item Bits can take on any value, while Qubits can only take on 0 and 1.
    \item \textbf{Bits can only take on the values 0 and 1, while Qubits can take on both 0 and 1 simultaneously.}
    \item Bits can only take the values 0 and 1, while Qubits can take the values 0, 1, and -1.
\end{enumerate}

\noindent What is the state of a Qubit before measurement?
\begin{enumerate}[label=\alph*)] \setlength{\itemsep}{0pt}
    \item A Qubit can only be in one definite state. 
    \item \textbf{A qubit can be in multiple states.}
    \item A Qubit can only be in two states simultaneously.
    \item A Qubit constantly switches between different states. 
\end{enumerate}

\noindent What are Qubits used for, among other things?
\begin{enumerate}[label=\alph*)] \setlength{\itemsep}{0pt}
    \item For storing and processing classical data in computers.
    \item To determine the energy of light.
    \item For generating light in optical devices.
    \item \textbf{For storing and processing information in quantum computers.}
\end{enumerate}

\noindent What is the Bloch sphere used for?
\begin{enumerate}[label=\alph*)] \setlength{\itemsep}{0pt}
    \item To represent the energy states of an atom.
    \item \textbf{To visualize the states of a Qubit.}
    \item To visualize the states of a classical Bit.
    \item To illustrate changes in the state of a laser.
\end{enumerate}

\noindent Based on the following representation of the Bloch sphere, what is the probability of measuring the Qubit in the state $\ket{0}$ or $\ket{1}$?

\begin{minipage}[t]{0.3\linewidth}
    \strut\vspace*{-\baselineskip}\newline\newline\includegraphics[width=\linewidth]{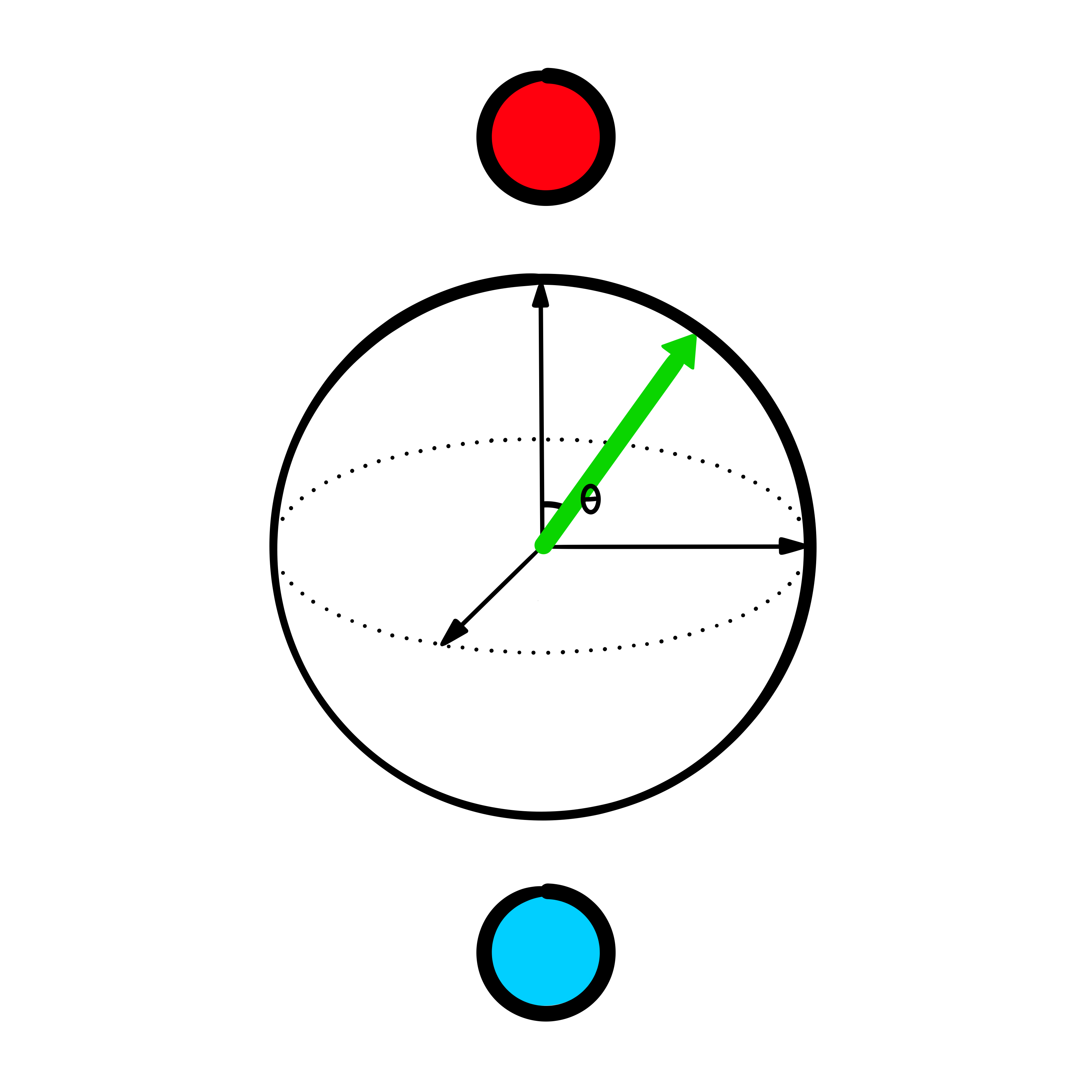}
    \label{fig:placeholder}
\end{minipage}
\begin{minipage}[t]{0.6\linewidth}
\vspace{0.2cm}
\begin{enumerate}[label=\alph*)] \setlength{\itemsep}{0pt}
    \item 40\% red, 60\% blue
    \item 60\% red, 40\% blue
    \item 10\% red, 90\% blue
    \item \textbf{90\% red, 10\% blue}
\end{enumerate}
\end{minipage}

\noindent Where does the arrow of the Bloch sphere point when the qubit is in a base state?

\begin{minipage}[t]{0.3\linewidth}
    \strut\vspace*{-\baselineskip}\newline\newline\includegraphics[width=\linewidth]{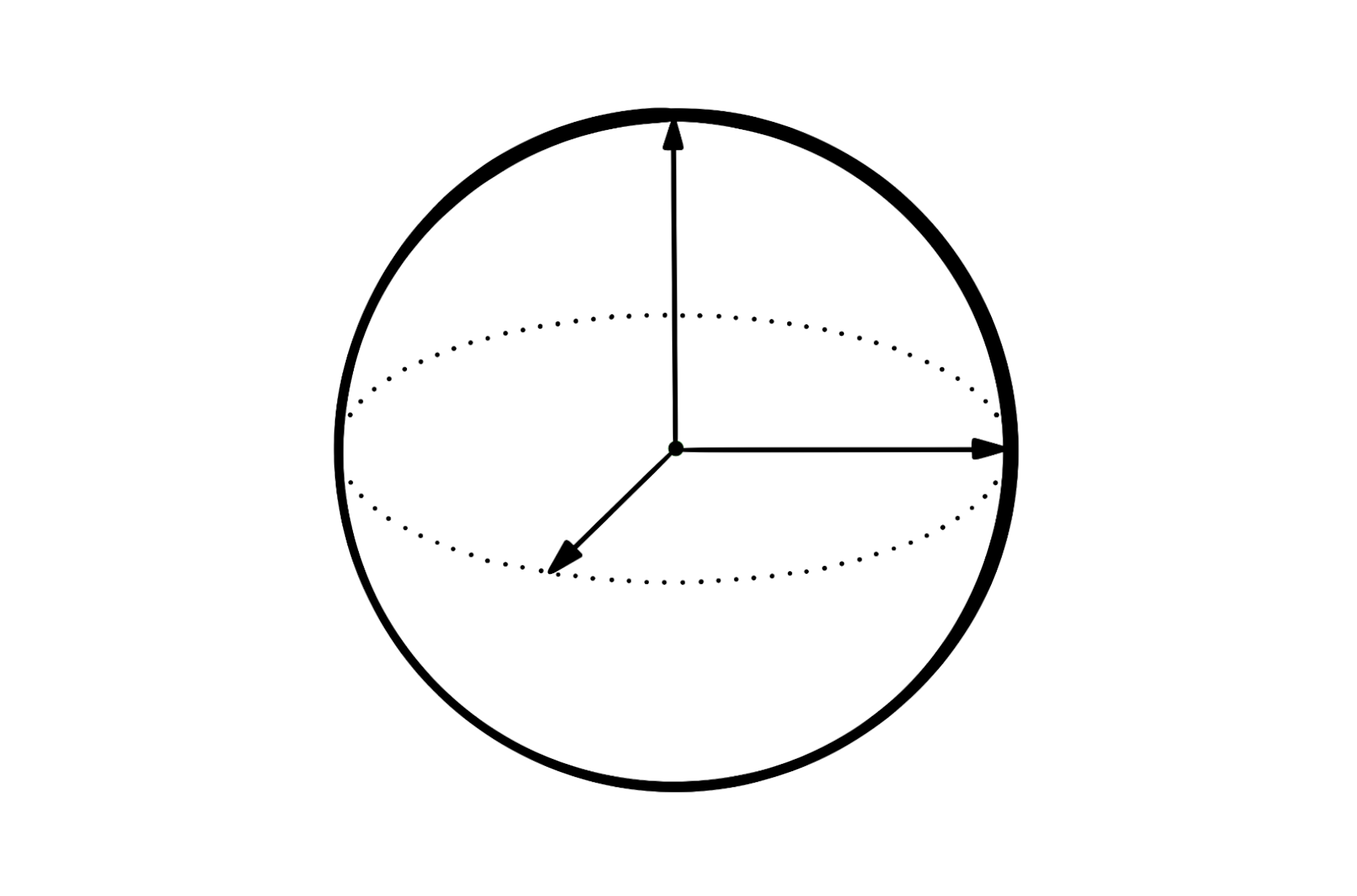}
    \label{fig:placeholder}
\end{minipage}
\begin{minipage}[t]{0.6\linewidth}
\vspace{0.2cm}
\begin{enumerate}[label=\alph*)] \setlength{\itemsep}{0pt}
    \item To the North Pole.
    \item To the South Pole.
    \item \textbf{To one of both Poles.}
    \item Anywhere but one of the Poles.
\end{enumerate}
\end{minipage}

\noindent Which of the following statements best describes the principle of superposition in quantum physics?
\begin{enumerate}[label=\alph*)] \setlength{\itemsep}{0pt}
    \item A quantum system can only be in one specific state.
    \item \textbf{A quantum system can exist in multiple states simultaneously.}
    \item A quantum system cannot change its state.
    \item A quantum system is always in a defined state.
\end{enumerate}

\noindent How does superposition in quantum mechanics differ from classical probability? 
\begin{enumerate}[label=\alph*)] \setlength{\itemsep}{0pt}
    \item \textbf{Superposition describes the simultaneous existence of multiple states, whereas classical probability is only a prediction of possible outcomes.}
    \item Superposition is only a mathematical construct, whereas classical probability describes real observable phenomena.
    \item Superposition and classical probability are identical concepts and do not differ from each other.
    \item Superposition only occurs with very large objects, whereas classical probability can only be applied to very small things, such as atoms.
\end{enumerate}

\noindent What does the expression “the Qubit collapses” mean?
\begin{enumerate}[label=\alph*)]
    \item The Qubit is destroyed by an external force.
    \item The Qubit changes its color.
    \item \textbf{The Qubit takes on a specific state when it is measured after being in superposition.}
    \item The Qubit becomes invisible.
\end{enumerate}

\noindent A Qubit is in a superposition. What happens to the qubit when a measurement is performed?
\begin{enumerate}[label=\alph*)] \setlength{\itemsep}{0pt}
    \item It stays in superposition.
    \item \textbf{It collapses into one of the base states.}
    \item It disappears.
    \item It splits into two Qubits
\end{enumerate}

\noindent Based on the following representation of the Bloch sphere, what is the probability of measuring the qubit in the “red” or “blue” state?

\begin{minipage}[t]{0.3\linewidth}
    \strut\vspace*{-\baselineskip}\newline\newline\includegraphics[width=\linewidth]{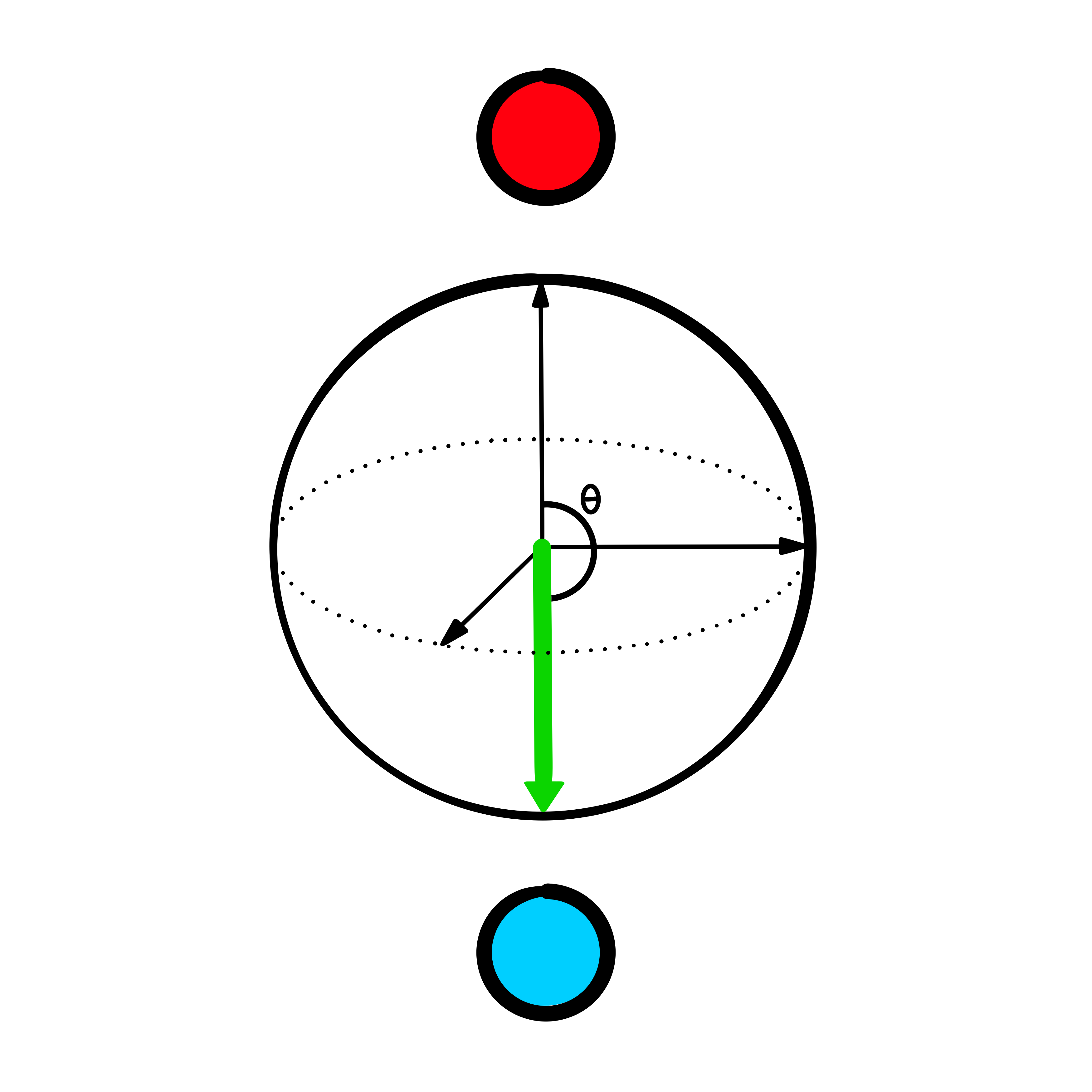}
    \label{fig:placeholder}
\end{minipage}
\begin{minipage}[t]{0.6\linewidth}
\vspace{0.2cm}
\begin{enumerate}[label=\alph*)] \setlength{\itemsep}{0pt}
    \item 100\% red
    \item \textbf{100\% blue}
    \item 50\% red, 50\% blue
    \item None of the above.
\end{enumerate}
\end{minipage}

\FloatBarrier

\noindent The probability of measuring the Qubit in the “red” state is 50\%.
Select the correct representation of the state on the Bloch sphere.

\begin{center}
\begin{minipage}[t]{0.24\columnwidth}
  \centering
  \includegraphics[width=\linewidth]{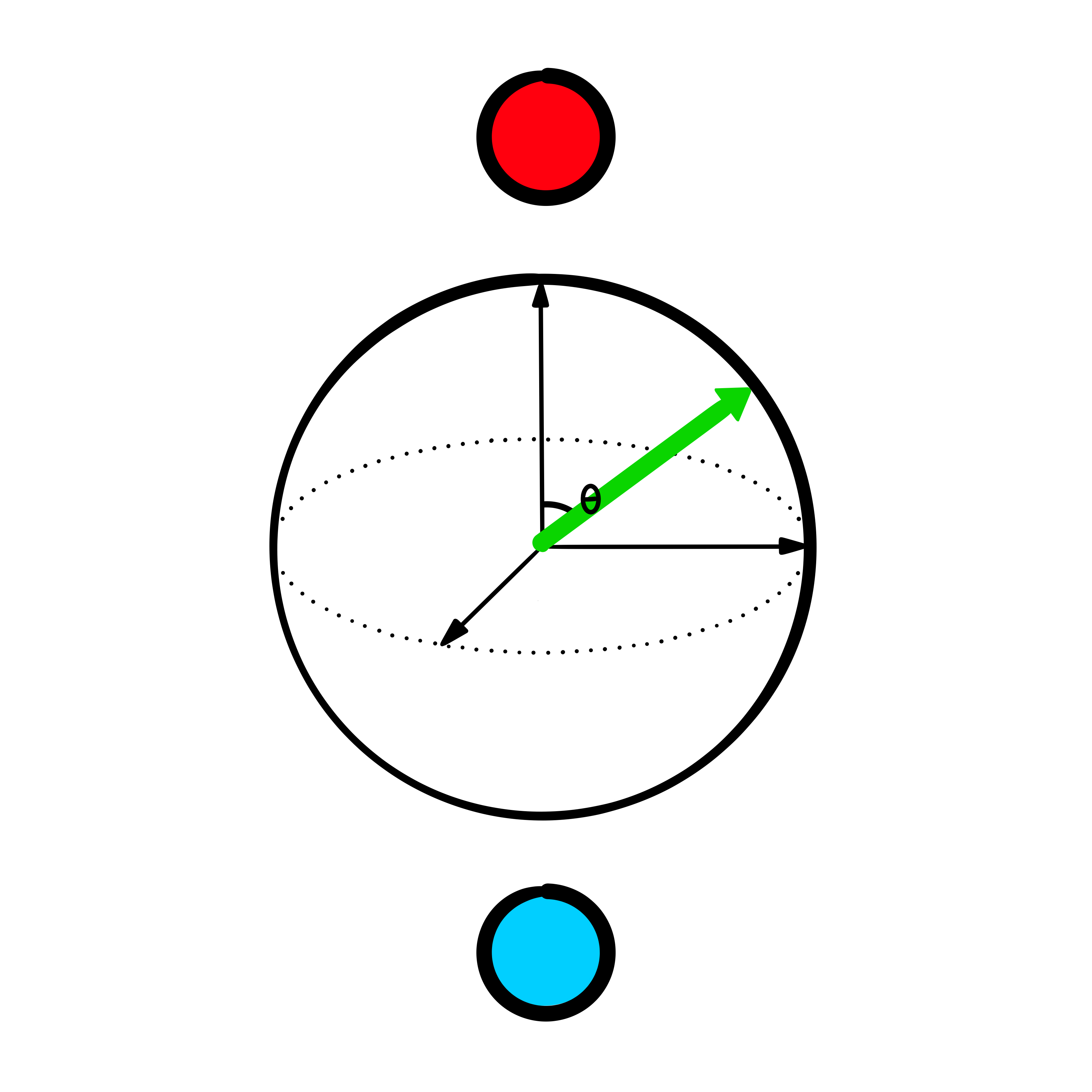}\\[-2pt]
  a)
\end{minipage}\hfill
\begin{minipage}[t]{0.24\columnwidth}
  \centering
  \includegraphics[width=\linewidth]{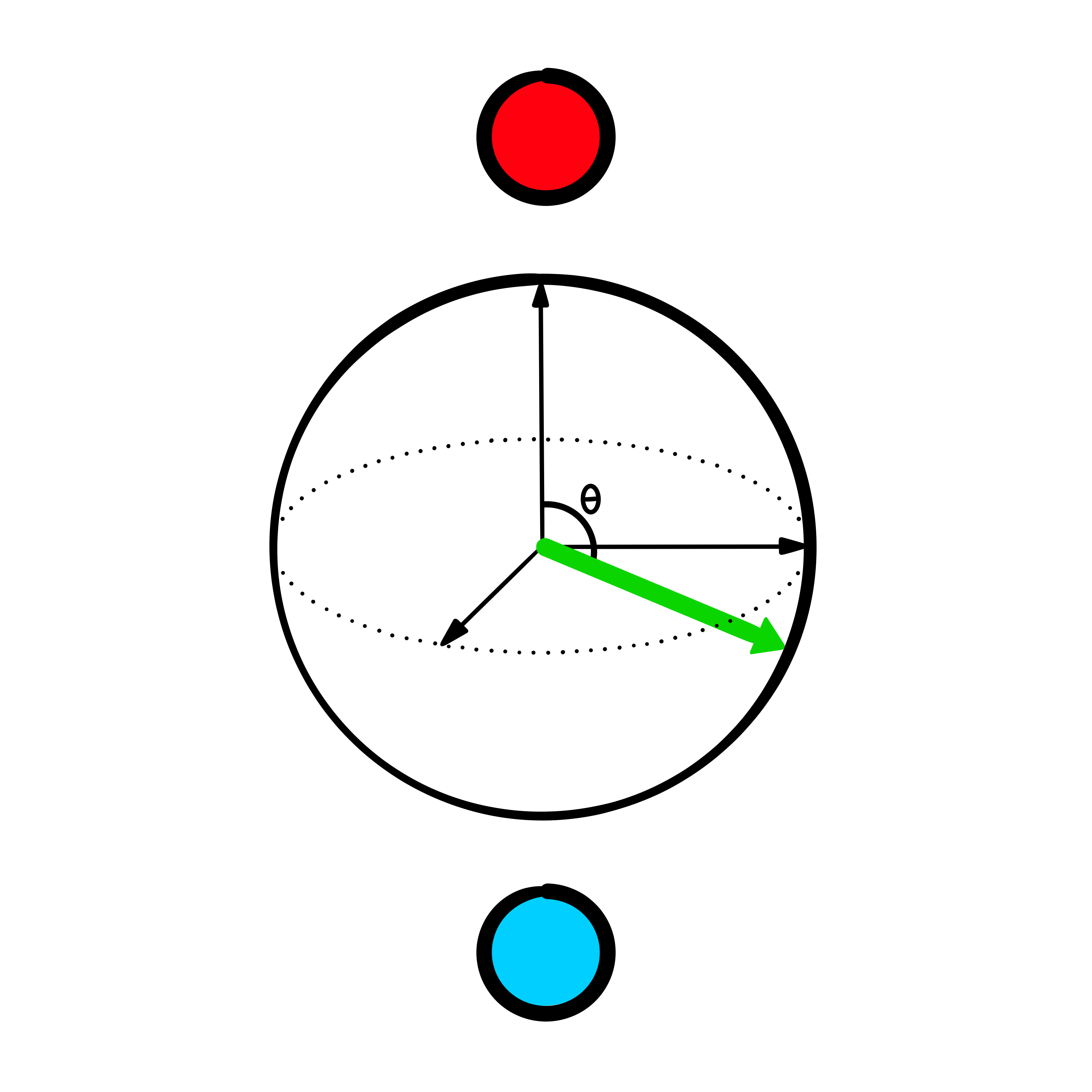}\\[-2pt]
  b)
\end{minipage}\hfill
\begin{minipage}[t]{0.24\columnwidth}
  \centering
  \includegraphics[width=\linewidth]{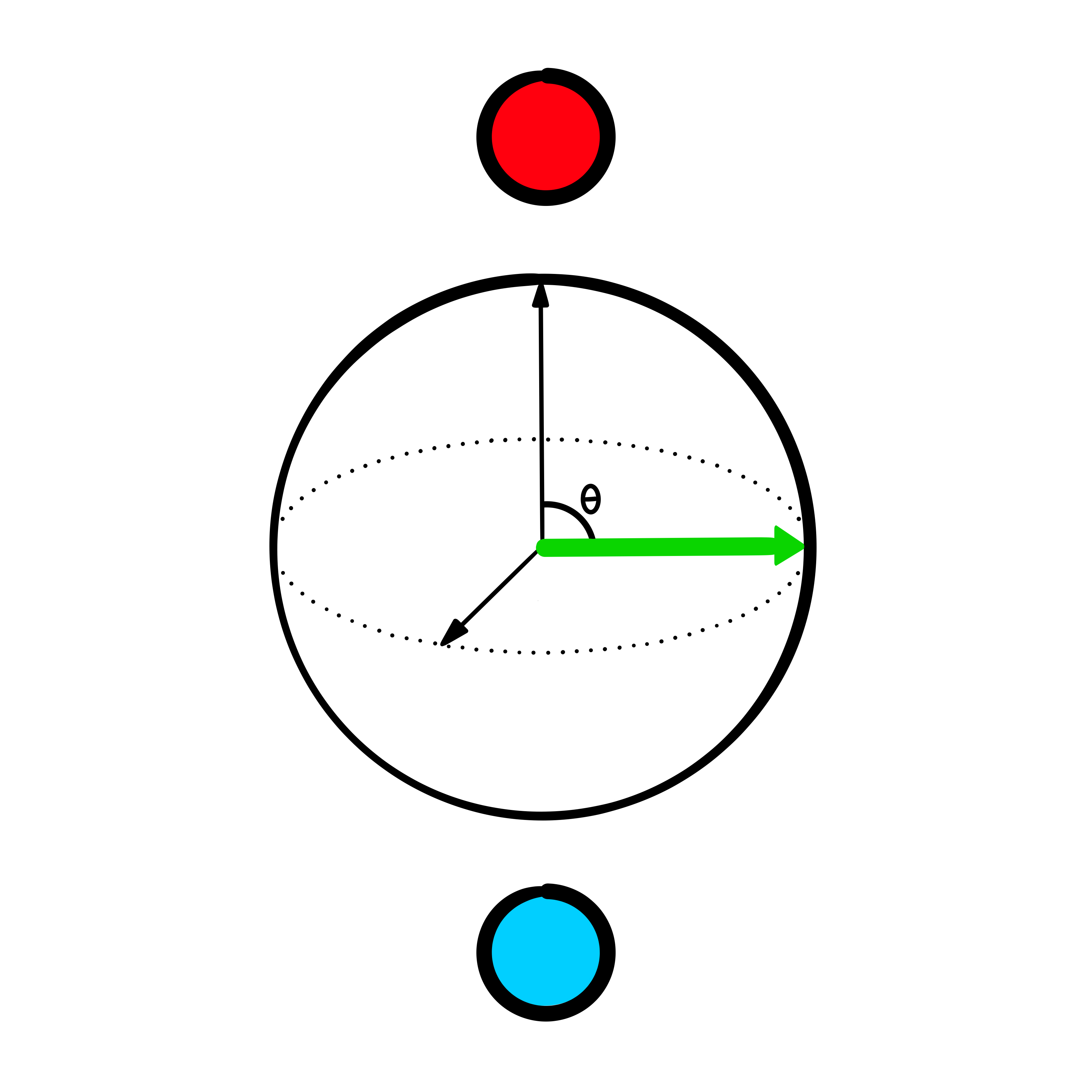}\\[-2pt]
  c)*
\end{minipage}\hfill
\begin{minipage}[t]{0.24\columnwidth}
  \centering
  \includegraphics[width=\linewidth]{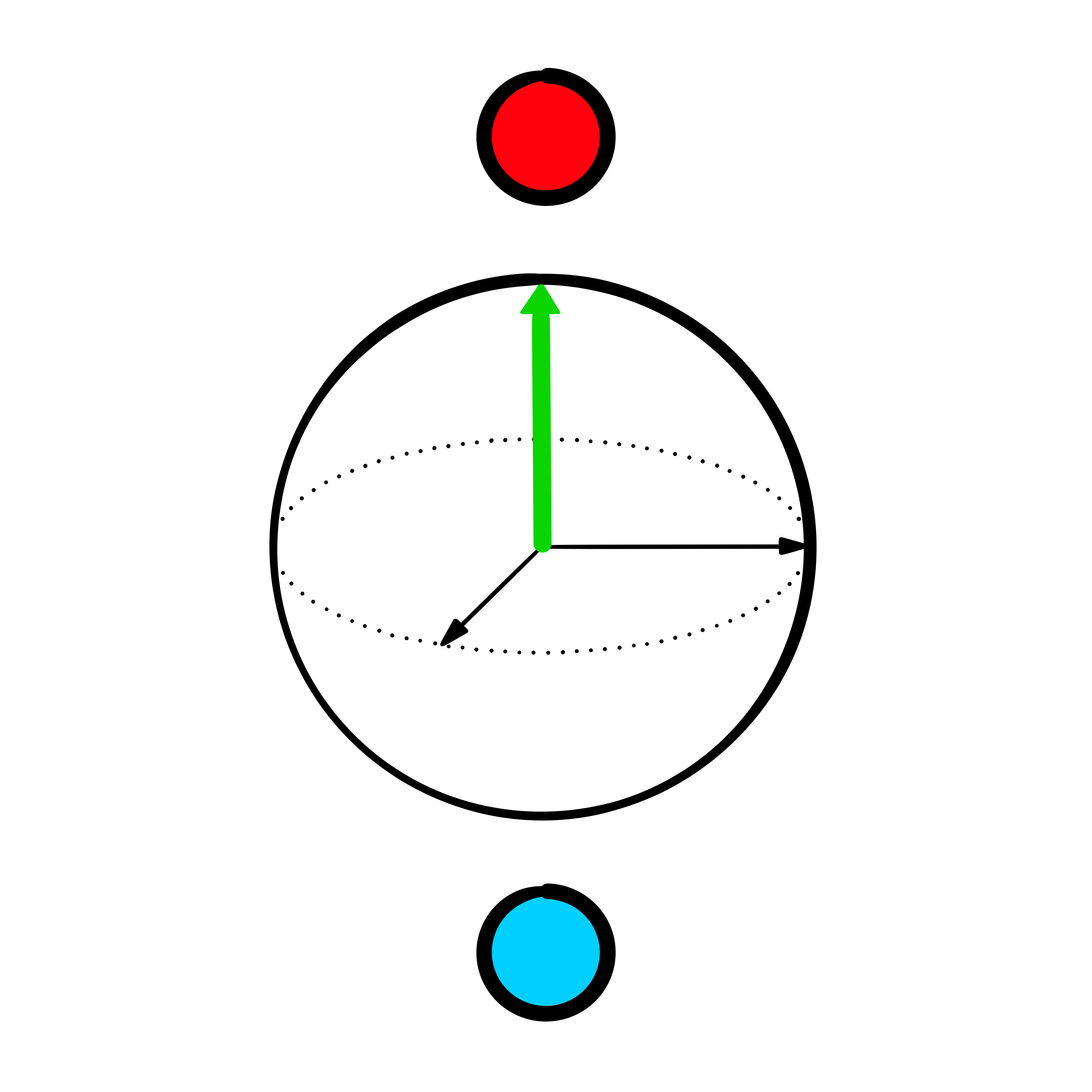}\\[-2pt]
  d)
\end{minipage}
\end{center}

\noindent The probability of measuring the Qubit in the “blue” state is 20\%.
Select the correct representation of the state on the Bloch sphere.

\begin{center}
\begin{minipage}[t]{0.24\columnwidth}
  \centering
  \includegraphics[width=\linewidth]{figs/54.jpg}\\[-2pt]
  a)*
\end{minipage}\hfill
\begin{minipage}[t]{0.24\columnwidth}
  \centering
  \includegraphics[width=\linewidth]{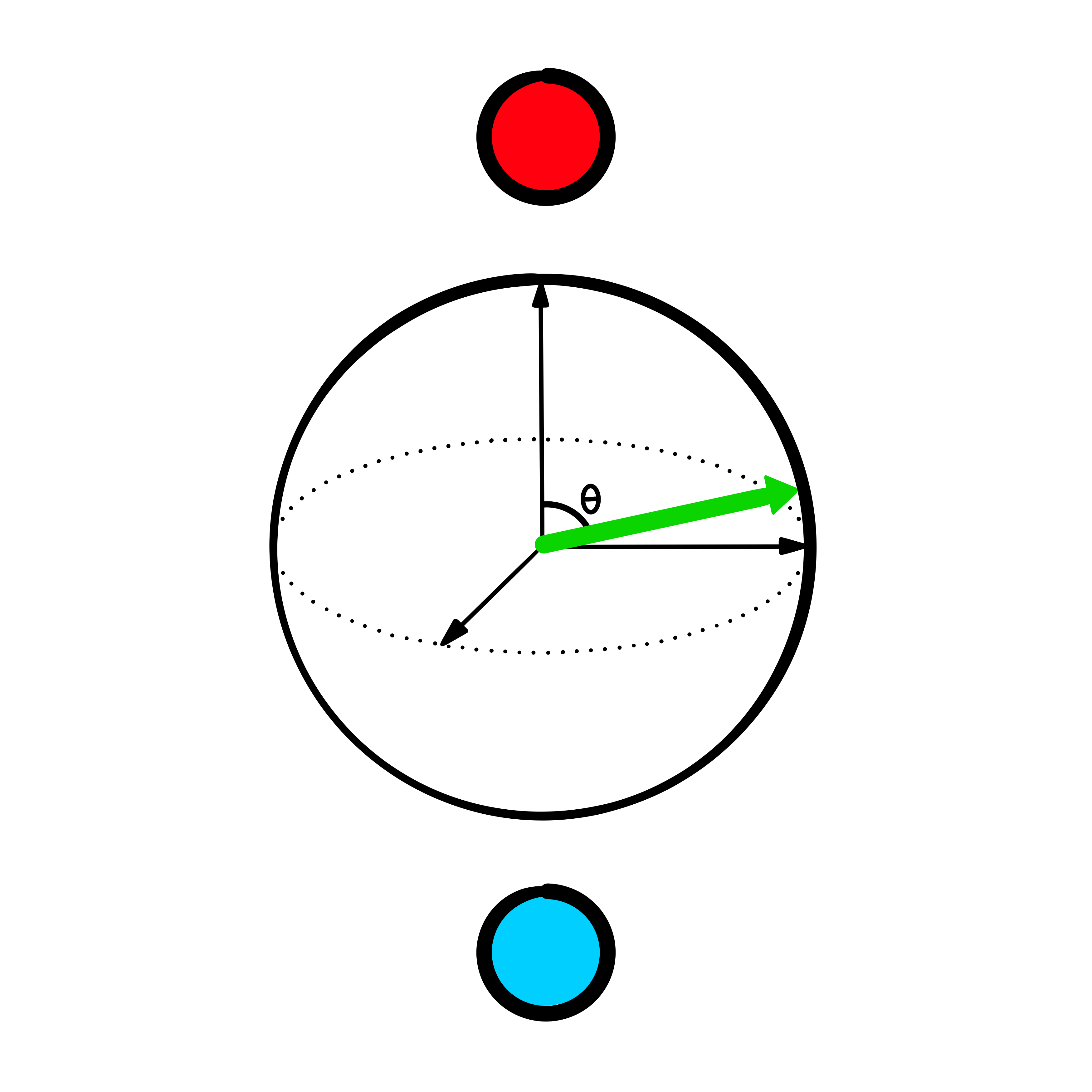}\\[-2pt]
  b)
\end{minipage}\hfill
\begin{minipage}[t]{0.24\columnwidth}
  \centering
  \includegraphics[width=\linewidth]{figs/113.png}\\[-2pt]
  c)
\end{minipage}\hfill
\begin{minipage}[t]{0.24\columnwidth}
  \centering
  \includegraphics[width=\linewidth]{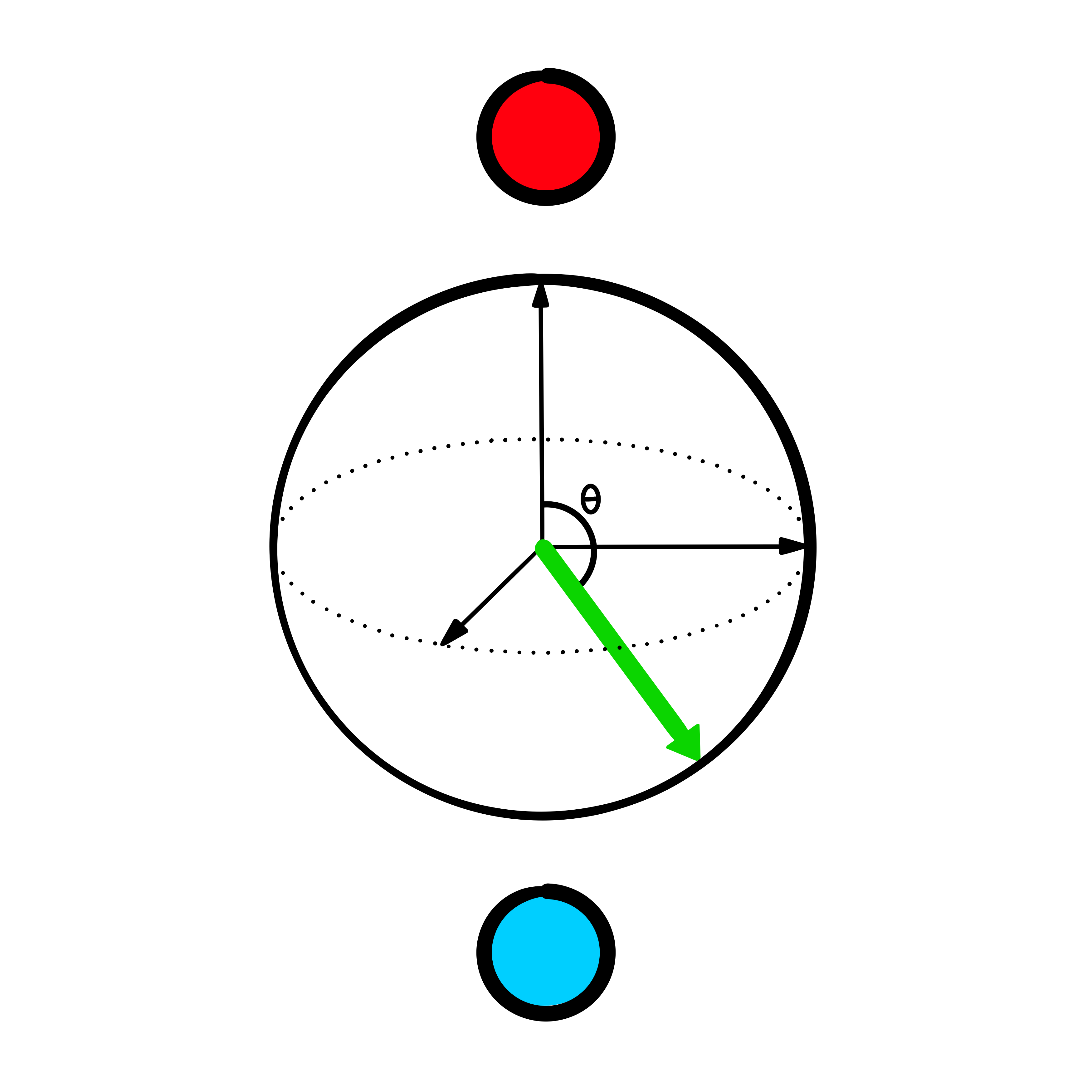}\\[-2pt]
  d)
\end{minipage}
\end{center}

\FloatBarrier

\section{Cognitive Load Questionnaire}
\label{app:CL_questionnaire}
\FloatBarrier

Cognitive load was assessed using nine items adapted from \cite{leppink2013}. The items were contextualized to the serious game environment used in this study. Participants rated each item on a 7-point Likert scale ranging from 0 (\textit{not at all the case}) to 6 (\textit{completely the case}).

The questionnaire includes items measuring intrinsic cognitive load (ICL), extraneous cognitive load (ECL), and germane cognitive load (GCL), as shown in Table~\ref{tab:CL_items}.

\begin{table}[h!]
\centering
\begin{tabularx}{\hsize}{lX}
\textbf{Type} & \textbf{Item wording} \\
\toprule
\addlinespace[0.5em]

ICL & The topic(s) covered in the game were very complex. \\
ICL & The game covered concepts and definitions that I perceived as very complex. \\
\addlinespace[0.5em]
\midrule
\addlinespace[0.5em]
ECL & The instructions and/or explanations were, in terms of learning, very ineffective. \\
ECL & The instructions and/or explanations during the game were very unclear. \\
ECL & The instructions and/or explanations were expressed using unclear language. \\
\addlinespace[0.5em]
\midrule
\addlinespace[0.5em]
GCL & The game enhanced my understanding of the topic(s) covered. \\
GCL & The game enhanced my knowledge and understanding of quantum physical phenomena and relations. \\
GCL & The game enhanced my understanding of the representations used. \\
GCL & The game enhanced my understanding of concepts and definitions. \\
\addlinespace[0.5em]

\bottomrule
\end{tabularx}
\caption{Items used to assess intrinsic (ICL), extraneous (ECL), and germane cognitive load (GCL), adapted from \cite{leppink2013}.}
\label{tab:CL_items}
\end{table}

\FloatBarrier


\begingroup
\setlength{\bibsep}{0pt plus 0.1ex}
\bibliographystyle{apsrev4-2}
\bibliography{mybib}
\endgroup

%\bibliographystyle{apalike} 
%\bibliography{mybib}



\end{document}